\documentclass[final,5p,times,twocolumn]{elsarticle}
\usepackage{etex}
\usepackage{lineno,hyperref}
\usepackage{amsmath}
\usepackage{amsfonts}
\modulolinenumbers[5]
\usepackage{array}
\usepackage{subfig}
\usepackage{gensymb}
\usepackage{cancel}
\usepackage[linesnumbered,ruled,vlined]{algorithm2e}% http://ctan.org/pkg/algorithm2e
\usepackage{morefloats}
\usepackage{mathrsfs}
\usepackage{relsize}

\usepackage{mathtools}
\usepackage{xcolor}
\usepackage[framemethod=tikz]{mdframed}
\usepackage{cancel}
\usepackage{breqn}
\DeclareMathAlphabet\mathbfcal{OMS}{cmsy}{b}{n}
\usepackage[many]{tcolorbox}
\newtcolorbox{highlighted}{colback=yellow,coltext=black,breakable}
\usepackage{todonotes}

\begin{document}

\begin{frontmatter}

\title{A novel study of temperature effects on the viscoelastic behavior of articular cartilage}

\author[JHU]{Reza Behrou\corref{cor1}}
\ead{rbehrou@jhu.edu}

\author[JHU]{Hamid Foroughi}

\author[JHU]{Fardad Haghpanah}

\cortext[cor1]{Corresponding Author: Reza Behrou}

\address[JHU]{Department of Civil Engineering, Johns Hopkins University, Baltimore, MD, USA}

%================================================================================
%
% Abstract
\begin{abstract}
This paper presents a new approach to study the effects of temperature on the poro- elastic and viscoelastic behavior of articular cartilage. Biphasic solid-fluid mixture theory is applied to study the poro-mechancial behavior of articular cartilage in a fully saturated state. The balance of linear momentum, mass, and energy are considered to describe deformation of the solid skeleton, pore fluid pressure, and temperature distribution in the mixture. The mechanical model assumes both linear elastic and viscoelastic isotropic materials, infinitesimal strain theory, and a time-dependent response. The influence of temperature on the mixture behavior is modeled through temperature dependent mass density and volumetric thermal strain. The fluid flow through the porous medium is described by the Darcy's law. The stress-strain relation for time-dependent viscoelastic deformation in the solid skeleton is described using the generalized Maxwell model. A verification example is presented to illustrate accuracy and efficiency of the developed finite element model. The influence of temperature is studied through examining the behavior of articular cartilage for confined and unconfined boundary conditions. Furthermore, articular cartilage under partial loading condition is modeled to investigate the deformation, pore fluid pressure, and temperature dissipation processes. The results suggest significant impacts of temperature on both poro- elastic and viscoelastic behavior of articular cartilage.
\end{abstract}


\end{frontmatter}
%

%================================================================================
%
% Introduction
\section{Introduction
\label{sec:Introduction}}
\begin{figure}[!h]
\centering
  \includegraphics[width=0.72\linewidth]{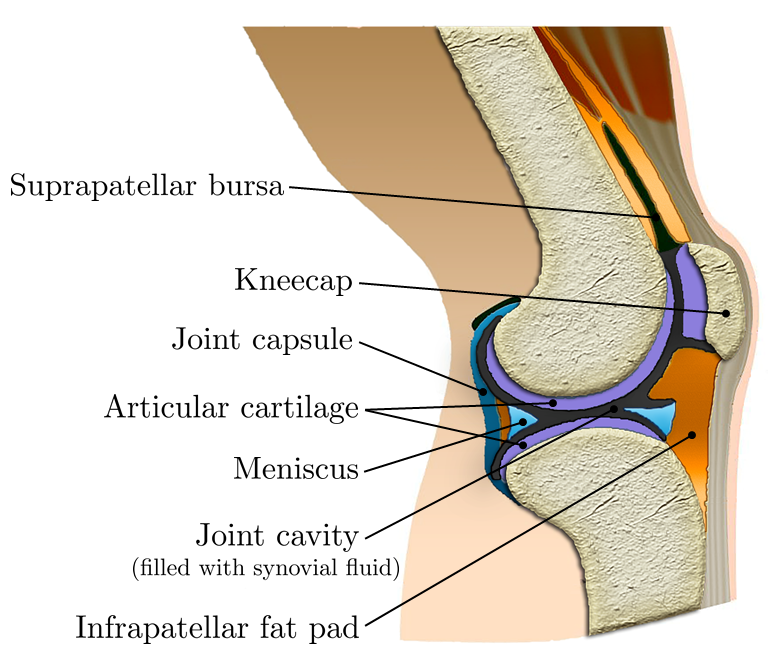}
\caption{Schematic diagram of the knee joint.}
\label{fig:ArticularCartilageSchematic}
\end{figure}

Diarthodial joints are the most common movable skeletal joints which are specifically characterized by some common structural features including: a layer of fibrocartilage that covers the opposing bony surfaces; a lubricating synovial fluid within the joint cavity; and an enclosing fibrous capsule which is lined with an active tissue (i.e. the synovium) \cite{MAS:93}. Figure \ref{fig:ArticularCartilageSchematic} illustrates the structure of knee joint as an example of diarthodial joints. 

Articular cartilage is a thin connective fibrocartilage (Figure \ref{fig:ArticularCartilageSchematic} that primarily consists of water, collagens, and proteoglycans. With 80\% of the wet weight, water is the dominant constituent of articular cartilage. The network of collagen, which constitutes 60\% of the dry weight, together with water increase the strength of the tissue to transmit high instantaneous loads. For smooth joint articulation, articular cartilage provides a lubricated surface that facilitates load transmission to the underlying subchondral bone \cite{FBR:09}. 

The mechanical behavior of articular cartilage has been studied using different models. Pena et al. \cite{PCM+:05} studied the effect of meniscal tears and meniscectomies on knee joints considering cartilage as a linear elastic isotropic and homogeneous material. Viscoelastic models are widely used to study the influences of an interstitial fluid on soft biological tissues \cite{Mak:86, SSH:08, GLA+:09, GL:11}. Poro-elastic and poro-viscoelatic models are developed to account for fluid flow in articular cartilage where biphasic mixture theory is used to describe deformation of the solid and the solid-fluid interactions \cite{HSK+:03, NTJ:03, HA:09, TGF:14}.

While most current models have studied the poro- elastic and viscoelastic behavior of articular cartilage, experimental studies \cite{JF:10} revealed that increasing temperature affects the behavior of both fluid and solid phases. Increasing temperature increases equilibrium stiffness and stress relaxation at higher temperatures, and decreases fluid viscosity. A comprehensive understanding of articular cartilage mechanics requires considering the effects of temperature. This study aims to investigate the influence of temperature on the poro- elastic and viscoelastic behavior of articular cartilage.

In this paper, a finite element model is developed to explore the thermal effects coupled with the poro- elastic and viscoelastic behavior of articular cartilage. The poro-mechanical behavior of a fully saturated articular cartilage is modeled through the solid-fluid mixture theory. The mechanical deformation of the solid skeleton, fluid flow, and temperature distribution in the solid-fluid mixture are modeled by the balances of linear momentum, mass, and energy. The mechanical behavior of the solid skeleton is described by both linear elastic and viscoelastic isotropic materials, infinitesimal strain theory, and a transient response. The motion of fluid through the porous medium is described by the Darcy's law. The influence of temperature on the mixture behavior is modeled by temperature dependent mass density and volumetric thermal strain. The governing equations are discretized in time with an implicit backward Euler scheme and in space by a standard Galerkin finite element method. The accuracy and efficiency of the finite element model are evaluated using analytical reference solutions.

The remainder of the paper is organized as follows: in Section \ref{sec:Methodology}, the methodology for theoretical formulation of the thermo-poro- elastic and viscoelastic models is elaborated. The numerical implementation and finite element analysis are described in Section \ref{sec:FiniteElementAnalysis}. In Section \ref{sec:ResultsAndDiscussion}, the effects of material type, temperature, and boundary conditions on the behavior of articular cartilage are discussed. Insights gained from the results and areas for future research are summarized in Section \ref{sec:Conclusions}.
%

%================================================================================
%
% Methodology
\section{Methodology
\label{sec:Methodology}}
The proposed thermo-poro-mechanical model predicts the coupled thermo-poro- elastic and viscoelastic behavior of articular cartilage. The model describes the mechanical deformation, pore fluid pressure and thermal phenomena in a fully saturated articular cartilage. The displacement of solid skeleton is modeled by the balance of linear momentum. The flow of fluid though the porous medium is described by the Darcy's law. The temperature evolution in the solid-fluid mixture is modeled by the balance of energy. The model for the thermo-poro- elastic and viscoelastic behavior of the saturated articular cartilage is described below.

\subsection{Basic definition of mixture}
Consider a porous medium consisting of a porous solid phase, denoted by $s$, with pore spaces occupied by a fluid, denoted by $\mathrm{f}$. The medium, therefore, is characterized by two phases. At time $t$, the current position of an $\alpha$ phase is defined by $\boldsymbol{x}^{\alpha}$ where $\alpha$ represents either the solid or the fluid phase. Throughout the paper, any variable appearing in the form of $(\bullet)^{\alpha}$ represents the $(\bullet)$ property of the $\alpha$ phase. The volumetric fraction occupied by the $\alpha$ phase is defined by:
\begin{equation}\label{eq:VolumeFraction_1}
\begin{aligned}
n^{\alpha}(\boldsymbol{x}^{\alpha},t) = \frac{dv^{\alpha}}{dv}, \qquad dv = J^{\alpha} dV^{\alpha}, \qquad J^{\alpha} = \det \boldsymbol{F}^{\alpha},
\end{aligned}
\end{equation}
where $dv^{\alpha}$ is the current differential volume, $dv$ is the current total differential volume of the mixture, $J^{\alpha}$ is the Jacobian of the deformation matrix, and $\boldsymbol{F}^{\alpha}$ is the deformation gradient. In biphasic mixture theory, $n^{s} + n^{\mathrm{f}} = 1$ where $n^{s}$ and $n^{\mathrm{f}}$ are volume fractions of the solid and fluid phases, respectively. The real ($\rho^{\alpha R}$) and partial ($\rho^{\alpha}$) mass densities of the $\alpha$ phase and total mass density, $\rho$, are defined as follows:
\begin{equation}\label{eq:AllMassDensity}
\begin{split}
\rho^{\alpha R}(\boldsymbol{x}^{\alpha},t) &= \frac{dm^{\alpha}}{dv^{\alpha}} \\
\rho^{\alpha}(\boldsymbol{x}^{\alpha},t) &= \frac{dm^{\alpha}}{dv} = \rho^{\alpha R}(\boldsymbol{x}^{\alpha},t) \ n^{\alpha}(\boldsymbol{x}^{\alpha},t) \\
\rho (\boldsymbol{x}^{\alpha},t) &= \rho^{s}(\boldsymbol{x}^{\alpha},t) + \rho^{\mathrm{f}}(\boldsymbol{x}^{\alpha},t)
\end{split}
,
\end{equation}
where $dm^{\alpha}$ is the differential mass. 

\subsection{ Balance of mass}
For the $\alpha$ phase, the balance of mass is defined as follows:
\begin{equation}\label{eq:BalaceofMassofalpha_3}
\begin{aligned}
\frac{D^{\alpha} \rho^{\alpha}}{Dt} + \rho^{\alpha} \nabla \cdot \boldsymbol{v}^{\alpha} = \gamma^{\alpha},
\end{aligned}
\end{equation}
where $\gamma^{\alpha}$ is the mass supply and $\nabla \cdot (\bullet) = \frac{\partial (\bullet)_{i}}{\partial x_{i}}$ is the divergence of a field. Considering volumetric thermal expansion for the $\alpha$ phase, the temperature dependent real mass density is expressed as follows \cite{LS:99}:
\begin{equation}\label{eq:BalaceofMassofalpha_4}
\begin{aligned}
\frac{1}{\rho^{\alpha R}}~\frac{D^{\alpha} \rho^{\alpha R}}{Dt} = \frac{1}{K^{\alpha}}\frac{D^{\alpha} p^{\alpha}}{Dt} - 3 \bar \alpha^{\alpha}_{\theta} \frac{D^{\alpha} \theta^{\alpha}}{Dt},
\end{aligned}
\end{equation}
where $K^{\alpha}$ is the bulk modulus, $p^{\alpha}$ is the pressure, $\bar \alpha^{\alpha}_{\theta}$ is the thermal expansion coefficient, and $\theta^{\alpha}$ is the temperature field. From  material time derivative of the fluid motion with respect to the motion of the solid phase, the following equations are derived \cite{Deboer:06}:
\begin{equation}\label{eq:BalaceofMassofalphaSolid1}
\begin{split}
\frac{D^{\mathrm{f}} n^{\mathrm{f}}}{Dt} = \frac{D^{s} n^{\mathrm{f}}}{Dt} + \nabla n^{\mathrm{f}} \cdot \boldsymbol{\tilde v}^{\mathrm{f}} \\
\frac{D^{\mathrm{f}} p^{\mathrm{f}}}{Dt} = \frac{D^{s} p^{\mathrm{f}}}{Dt} + \nabla p^{\mathrm{f}} \cdot \boldsymbol{\tilde v}^{\mathrm{f}} \\
\frac{D^{\mathrm{f}} \theta^{\mathrm{f}}}{Dt} = \frac{D^{s} \theta^{\mathrm{f}}}{Dt} + \nabla \theta^{\mathrm{f}} \cdot \boldsymbol{\tilde v}^{\mathrm{f}}
\end{split}
,
\end{equation}
where $\boldsymbol{\tilde v}^{\mathrm{f}} = \boldsymbol{v}^{\mathrm{f}} - \boldsymbol{v}^{s}$ is the relative velocity of the fluid phase with respect to the motion of the solid phase. Use of Eqs. \eqref{eq:AllMassDensity}, \eqref{eq:BalaceofMassofalpha_4}, and \eqref{eq:BalaceofMassofalphaSolid1} in \eqref{eq:BalaceofMassofalpha_3}, the balance of mass for the mixture, with respect to the solid phase motion, which accounts for the volumetric thermal expansion in both solid and fluid phases, is expressed as follows:
\begin{equation}\label{eq:BalaceofMassofalpha_10}
\begin{split}
\frac{n^{s}}{K^{s}}~\frac{D^{s} p^{s}}{Dt} + \frac{n^{\mathrm{f}}}{K^{\mathrm{f}}}~\frac{D^{s} p^{\mathrm{f}}}{Dt} + \frac{1}{K^{\mathrm{f}}} \ \nabla p^{\mathrm{f}} \cdot (n^{\mathrm{f}}~\boldsymbol{\tilde v}^{\mathrm{f}}) \\
- 3 n^{s} \bar \alpha^{s}_{\theta} \frac{D^{s} \theta^{s}}{Dt} - 3 n^{\mathrm{f}} \bar \alpha^{\mathrm{f}}_{\theta} \left ( \frac{D^{s} \theta^{\mathrm{f}}}{Dt} +  \nabla \theta^{\mathrm{f}} \cdot \boldsymbol{\tilde v}^{\mathrm{f}} \right ) \\
+ ~\nabla \cdot \boldsymbol{v}^{s} + \nabla \cdot (n^{\mathrm{f}} \boldsymbol{\tilde v}^{\mathrm{f}}) = \frac{\gamma^{s}}{\rho^{s R}} + \frac{\gamma^{\mathrm{f}}}{\rho^{\mathrm{f} R}}.
\end{split}
\end{equation}

\subsection{Balance of linear momentum}
The balance of linear momentum for the $\alpha$ phase is defined as:
\begin{equation}\label{eq:BalanceofLinearMomentum_11}
\begin{aligned}
\nabla \cdot \boldsymbol{\sigma}^{\alpha} + \rho^{\alpha} \boldsymbol{b}^{\alpha} + \boldsymbol{h}^{\alpha} = \rho^{\alpha} \boldsymbol{a}^{\alpha} + \gamma^{\alpha}\boldsymbol{v}^{\alpha},
\end{aligned}
\end{equation}
where $\boldsymbol{\sigma}^{\alpha}$ is the partial stress, $\boldsymbol{b}^{\alpha}$ is the body force vector, and $\boldsymbol{a}^{\alpha}$ is the acceleration. The total stress is defined as $\boldsymbol{\sigma} = \boldsymbol{\sigma}^{s} + \boldsymbol{\sigma}^{\mathrm{f}}$. The internal body force due to drag, $\boldsymbol{h}^{\alpha}$, on the $\alpha$ constituent caused by the other constituents is given by:
\begin{equation}\label{eq:BalanceofLinearMomentum_DragForce}
\begin{aligned}
\sum_{\alpha = s, \mathrm{f}} \boldsymbol{h}^{\alpha} = \boldsymbol{h}^{s} + \boldsymbol{h}^{\mathrm{f}} = 0.
\end{aligned}
\end{equation}
By summing up the individual components of the balance of linear momentum for the solid and fluid phases, the balance of linear momentum for the mixture is expressed as:
\begin{equation}\label{eq:BalanceofLinearMomentum_Total}
\begin{aligned}
\nabla \cdot \boldsymbol{\sigma} + \rho^{s} \boldsymbol{b}^{s} + \rho^{\mathrm{f}} \boldsymbol{b}^{\mathrm{f}} = \rho^{s} \boldsymbol{a}^{s} + \rho^{\mathrm{f}} \boldsymbol{a}^{\mathrm{f}} + \gamma^{s} \boldsymbol{v}^{s} + \gamma^{\mathrm{f}} \boldsymbol{v}^{\mathrm{f}}.
\end{aligned}
\end{equation}
Furthermore, the balance of angular momentum states that the stresses are symmetric, i.e. $\boldsymbol{\sigma}^{\alpha} = (\boldsymbol{\sigma}^{\alpha})^{T}$.

\subsection{ Balance of energy}
The first and second laws of thermodynamics for the $\alpha$ phase are given by \cite{Deboer:06}: 
\begin{equation}\label{eq:BalanceofEnergy_11}
\begin{split}
\rho^{\alpha} \frac{D^{\alpha} e^{\alpha}}{Dt} - \boldsymbol{\sigma}^{\alpha} \colon \boldsymbol{\dot \varepsilon}^{\alpha} + \nabla \cdot \boldsymbol{q}^{\alpha} - \rho^{\alpha} r^{\alpha} \\
- \gamma^{\alpha}  \left ( \frac{1}{2} \boldsymbol{v}^{\alpha} \cdot \boldsymbol{v}^{\alpha} - e^{\alpha} \right ) + \boldsymbol{h}^{\alpha} \cdot \boldsymbol{v}^{\alpha} - \hat e^{\alpha} = 0,
\end{split}
\end{equation}
\begin{equation}\label{eq:EntropyInequalityPrinciple_4}
\begin{aligned}
\gamma^{\alpha} \eta^{\alpha} \theta^{\alpha} + \rho^{\alpha} \theta^{\alpha} \frac{D^{\alpha} \eta^{\alpha}}{Dt} - \rho^{\alpha} r^{\alpha} - \frac{1}{\theta^{\alpha}} \nabla \theta^{\alpha} \cdot \boldsymbol{q}^{\alpha} + \nabla \cdot \boldsymbol{q}^{\alpha} \geq 0,
\end{aligned}
\end{equation}
where $e^{\alpha}$ is the internal energy per unit mass, $\boldsymbol{q}^{\alpha}$ is the heat flux vector, $r^{\alpha}$ is the heat input rate per unit mass, $\hat e^{\alpha}$ is the power density supply by other phases on the $\alpha$ phase, and $\eta^{\alpha}$ is the entropy per unit mass. The Helmholtz free energy per unit mass is defined as \cite{Deboer:06}:
\begin{equation}\label{eq:HelmHoltzFreeEnergy_1}
\begin{aligned}
\psi^{\alpha} = e^{\alpha} - \theta^{\alpha} \eta^{\alpha}.
\end{aligned}
\end{equation}
Use of the material time derivative of the Helmholtz free energy and the first law of thermodynamics in the second law of thermodynamics leads to the Clausius-Duhem inequality for the $\alpha$ phase as follows:
\begin{equation}\label{eq:HelmHoltzFreeEnergy_3}
\begin{aligned}
\gamma^{\alpha}  \left ( \eta^{\alpha} \theta^{\alpha} + \frac{1}{2} \boldsymbol{v}^{\alpha} \cdot \boldsymbol{v}^{\alpha} - e^{\alpha} \right ) + \boldsymbol{\sigma}^{\alpha} \colon \boldsymbol{\dot \varepsilon}^{\alpha} - \boldsymbol{h}^{\alpha} \cdot \boldsymbol{v}^{\alpha} + \hat e^{\alpha} \\
- \rho^{\alpha} \frac{D^{\alpha} \psi^{\alpha}}{Dt} - \rho^{\alpha} \eta^{\alpha} \frac{D^{\alpha} \theta^{\alpha}}{Dt} - \frac{1}{\theta^{\alpha}} \nabla \theta^{\alpha} \cdot \boldsymbol{q}^{\alpha} \geq 0.
\end{aligned}
\end{equation}

In the mixture theory, the total Cauchy stress, $\boldsymbol{\sigma}$, is defined as a function of the effective Cauchy stress, $\boldsymbol{\sigma}'$, acting on the solid skeleton, and the Cauchy pore fluid pressure, $\bar p^{\mathrm{f}}$:
\begin{equation}\label{eq:HelmHoltzFreeEnergy_4}
\begin{aligned}
\boldsymbol{\sigma} = \boldsymbol{\sigma}' - \tilde B \bar p^{\mathrm{f}} \boldsymbol{I},
\end{aligned}
\end{equation}
where $\tilde B = 1 - K_{\mathrm{skel}}/K^{s} \leq 1$ is the Biot coefficient, and $K_{\mathrm{skel}}$ and $K^{s}$ are the bulk moduli of the porous medium and solid phase, respectively \cite{GSG:96}. The fluid partial stress, $\boldsymbol{\sigma}^{\mathrm{f}}$, is described by fluid pressure and viscous terms as \cite{Holzapfel:00}:
\begin{equation}\label{eq:HelmHoltzFreeEnergy_5_1}
\begin{aligned}
\boldsymbol{\sigma}^{\mathrm{f}} = - p^{\mathrm{f}} \boldsymbol{I} + \boldsymbol{\sigma}^{\mathrm{f,visc}},
\end{aligned}
\end{equation}
where the partial fluid pressure is related to the Cauchy pore fluid pressure through $p^{\mathrm{f}} = n^{\mathrm{f}} \bar p^{\mathrm{f}}$. The fluid viscous stress is defined as \cite{Holzapfel:00}:
\begin{equation}\label{eq:HelmHoltzFreeEnergy_5_2}
\begin{aligned}
\boldsymbol{\sigma}^{\mathrm{f,visc}} = k^{\mathrm{f}} (\text{tr} \boldsymbol{d}^{\mathrm{f}}) \boldsymbol{I} + 2 \mu^{\mathrm{f}} \boldsymbol{d}^{\mathrm{f}}, \quad \boldsymbol{d}^{\mathrm{f}} = \frac{1}{2} \Big( \nabla \boldsymbol{v}^{\mathrm{f}} + (\nabla  \boldsymbol{v}^{\mathrm{f}})^{T} \Big),
\end{aligned}
\end{equation}
where $k^{\mathrm{f}}$ is the partial bulk viscosity, $\mu^{\mathrm{f}}$ is the partial shear viscosity, and $\boldsymbol{d}^{\mathrm{f}}$ is the symmetric portion of the velocity gradient for the fluid phase. Thus, the partial stress of the solid phase is derived as:
\begin{equation}\label{eq:HelmHoltzFreeEnergy_6}
\begin{split}
\boldsymbol{\sigma}^{s} = \boldsymbol{\sigma}' - (\tilde B - n^{\mathrm{f}}) \bar p^{\mathrm{f}} \boldsymbol{I} - \boldsymbol{\sigma}^{\mathrm{f,visc}}.
\end{split}
\end{equation}

For thermoelastic problems, the total strain tensor of the solid skeleton is decomposed into elastic and thermal components as follows:
\begin{equation}\label{eq:HelmHoltzFreeEnergy_11}
\begin{aligned}
\boldsymbol{\varepsilon}^{s} = \boldsymbol{\varepsilon}^{\mathrm{skel,e}} + \boldsymbol{\varepsilon}^{s,\mathrm{thm}}, \qquad \boldsymbol{\varepsilon}^{s,\mathrm{thm}} = \bar{\alpha}_{\theta}^{s} \big( \theta^{s} - \theta^{s}_{0} \big) \boldsymbol{I},
\end{aligned}
\end{equation}
where $\boldsymbol{\varepsilon}^{\mathrm{skel,e}}$ is the elastic mechanical strain, $\boldsymbol{\varepsilon}^{s,\mathrm{thm}}$ is the volumetric thermal strain, and $\theta^{s}_{0}$ is the initial temperature of the solid phase. Thus, the stress power for the solid-fluid mixture can be expressed as follows:
\begin{equation}\label{eq:HelmHoltzFreeEnergy_13_2}
\begin{split}
\boldsymbol{\sigma}^{s} \colon \frac{D^{s} \boldsymbol{\varepsilon}^{s}}{Dt} + \boldsymbol{\sigma}^{\mathrm{f}} \colon \frac{D^{\mathrm{f}} \boldsymbol{\varepsilon}^{\mathrm{f}}}{Dt} = \boldsymbol{\sigma'} \colon \frac{D^{s} \boldsymbol{\varepsilon}^{\mathrm{skel},e}}{Dt} + \bar{\alpha}_{\theta}^{s} \boldsymbol{\sigma'} \colon \boldsymbol{I} \frac{D^{s} \theta^{s}}{Dt} \\
- \tilde B \bar p^{\mathrm{f}} \nabla \cdot \boldsymbol{v}^{s} - n^{\mathrm{f}} \bar p^{\mathrm{f}} \nabla \cdot \boldsymbol{\tilde v}^{\mathrm{f}} + \boldsymbol{\sigma}^{\mathrm{f,visc}} \colon \nabla \boldsymbol{\tilde v}^{\mathrm{f}}.
\end{split}
\end{equation}

With the assumption that both solid and fluid constituents are nearly incompressible (i.e. $\rho^{\alpha R}$ is constant), the Helmholtz free energy per unit mass for the solid skeleton and the fluid phase is expressed as follows:
\begin{equation}\label{eq:HelmHoltzFreeEnergy_17}
\begin{aligned}
\psi^{s} = \psi^{s} (\boldsymbol{\varepsilon}^{\mathrm{skel},e}, \theta^{s}), \qquad  \psi^{\mathrm{f}} = \psi^{\mathrm{f}} (\theta^{\mathrm{f}}),
\end{aligned}
\end{equation}
with the following assumptions:
\begin{equation}\label{eq:HelmHoltzFreeEnergy_17_1}
\begin{aligned}
\frac{\partial^{2} \psi^{s}}{\partial \theta^{s} \partial \boldsymbol{\varepsilon}^{\mathrm{skel},e}} = 0, \qquad \frac{\partial^{2} \psi^{\alpha}}{\partial (\theta^{\alpha})^{2}} = -\frac{C_{p}^{\alpha}}{\theta^{\alpha}},
\end{aligned}
\end{equation}
where $C_{p}^{\alpha}$ is the specific heat capacity per unit mass. We further assume that (1) both solid and fluid constituents are nearly incompressible, i.e. $\frac{1}{K^{\alpha}} \approx 0$; (2) the fluid flow is inviscid and the viscous effects are neglected, i.e. $\boldsymbol{\sigma}^{\mathrm{f,visc}} = 0$; (3) there is no mass exchange between solid and fluid phases and the mass supply of the $\alpha$ phase is zero, i.e. $\gamma^{\alpha} = 0$; (4) the sum of the power of the solid and fluid phases is set to zero, i.e. $\hat e^{s} + \hat e^{\mathrm{f}} = 0$; (5) $\theta^{s} = \theta^{\mathrm{f}} = \theta$; (6) $\boldsymbol{q} = \boldsymbol{q}^{s} + \boldsymbol{q}^{\mathrm{f}}$; (7) $\boldsymbol{a}^{s} = \boldsymbol{a}^{\mathrm{f}} = 0$; (8) $\boldsymbol{b}^{s} = \boldsymbol{b}^{\mathrm{f}} = \boldsymbol{g}$; and (9) $\tilde B = 1$. Thus, the Clausius-Duhem inequality of the solid-fluid mixture is expressed as: 
\begin{equation}\label{eq:HelmHoltzFreeEnergy_20_1}
\begin{split}
\left( \boldsymbol{\sigma'} - \rho^{s} \frac{\partial \psi^{s}}{\partial \boldsymbol{\varepsilon}^{\mathrm{skel},e}} \right) \colon \frac{D^{s} \boldsymbol{\varepsilon}^{\mathrm{skel},e}}{Dt} \\
- \left( \rho^{s} \frac{\partial \psi^{s}}{\partial \theta} + \rho^{s} \eta^{s}  + 3 n^{s} \bar \alpha^{s}_{\theta} \bar p^{\mathrm{f}} - \bar{\alpha}_{\theta}^{s} \boldsymbol{\sigma'} \colon \boldsymbol{I} \right) \frac{D^{s} \theta}{Dt} \\
- \left( \rho^{\mathrm{f}} \frac{\partial \psi^{\mathrm{f}}}{\partial \theta} + \rho^{\mathrm{f}} \eta^{\mathrm{f}} + 3 n^{\mathrm{f}} \bar \alpha^{\mathrm{f}}_{\theta} \bar p^{\mathrm{f}} \right) \frac{D^{\mathrm{f}} \theta}{Dt} \\
- \left ( \nabla \bar p^{\mathrm{f}} - \rho^{\mathrm{fR}} \boldsymbol{g} \right ) \cdot n^{\mathrm{f}} \boldsymbol{\tilde v}^{\mathrm{f}} - \frac{1}{\theta} \nabla \theta \cdot \boldsymbol{q} \geq 0.
\end{split}
\end{equation}
The Inequalities in Eq. \eqref{eq:HelmHoltzFreeEnergy_20_1} must hold valid for all possible thermodynamic states. Following Coleman and Noll \cite{CN:63} argument for independent processes (i.e. $D^{s} \boldsymbol{\varepsilon}^{\mathrm{skel},e}/Dt$, $D^{s} \theta/Dt$, and , $D^{\mathrm{f}} \theta/Dt$), the following constitutive equations must hold:
\begin{equation}\label{eq:HelmHoltzFreeEnergy_21}
\begin{split}
\boldsymbol{\sigma'} = \rho^{s} \frac{\partial \psi^{s}}{\partial \boldsymbol{\varepsilon}^{\mathrm{skel},e}} \\
\rho^{s} \eta^{s} = -\rho^{s} \frac{\partial \psi^{s}}{\partial \theta} - 3 n^{s} \bar \alpha^{s}_{\theta} \bar p^{\mathrm{f}} + \bar{\alpha}_{\theta}^{s} \boldsymbol{\sigma'} \colon \boldsymbol{I} \\
\rho^{\mathrm{f}} \eta^{\mathrm{f}} = -\rho^{\mathrm{f}} \frac{\partial \psi^{\mathrm{f}}}{\partial \theta} - 3 n^{\mathrm{f}} \bar \alpha^{\mathrm{f}}_{\theta} \bar p^{\mathrm{f}}
\end{split}
.
\end{equation}
Thus, Eq. \eqref{eq:HelmHoltzFreeEnergy_20_1} reduces to:
\begin{equation}\label{eq:HelmHoltzFreeEnergy_24}
\begin{split}
- \left ( \nabla \bar p^{\mathrm{f}} - \rho^{\mathrm{fR}} \boldsymbol{g} \right ) \cdot \boldsymbol{\tilde v}^{\mathrm{f}}_{D} - \frac{1}{\theta} \nabla \theta \cdot \boldsymbol{q} \geq 0,
\end{split}
\end{equation}
where $\boldsymbol{\tilde v}^{\mathrm{f}}_{D} = n^{\mathrm{f}} \boldsymbol{\tilde v}^{\mathrm{f}}$ is the Darcy relative fluid velocity. In Eq. \eqref{eq:HelmHoltzFreeEnergy_24}, the constitutive equation for the Darcy relative fluid velocity gives non-negative dissipation \cite{Coussy:04}, where:
\begin{equation}\label{eq:HelmHoltzFreeEnergy_25}
\begin{split}
\boldsymbol{\tilde v}^{\mathrm{f}}_{D} = - k_{p} \left ( \nabla \bar p^{\mathrm{f}} - \rho^{\mathrm{fR}} \boldsymbol{g} \right ),
\end{split}
\end{equation}
and $k_{p}$ is the isotropic hydraulic permeability of the fluid. Finally, use of Eqs. \eqref{eq:HelmHoltzFreeEnergy_20_1} and \eqref{eq:HelmHoltzFreeEnergy_24} in \eqref{eq:BalanceofEnergy_11} leads to the balance of energy for the solid-fluid mixture as follows: 
\begin{equation}\label{eq:HelmHoltzFreeEnergy_33}
\begin{split}
\rho^{s} C_{p}^{s} \frac{D^{s} \theta}{Dt} - 3 n^{s} \bar \alpha^{s}_{\theta} \theta \frac{D^{s} \bar p^{\mathrm{f}}}{Dt} + \bar{\alpha}_{\theta}^{s} \theta \frac{D^{s} \boldsymbol{\sigma'}}{Dt} \colon \boldsymbol{I} \\
+ \rho^{\mathrm{f}} C_{p}^{\mathrm{f}} \Big( \frac{D^{s} \theta}{Dt} + \nabla \theta \cdot \boldsymbol{\tilde v}^{\mathrm{f}} \Big) - 3 n^{\mathrm{f}} \bar \alpha^{\mathrm{f}}_{\theta} \theta \Big( \frac{D^{s} \bar p^{\mathrm{f}}}{Dt} + \nabla \bar p^{\mathrm{f}} \cdot \boldsymbol{\tilde v}^{\mathrm{f}} \Big) \\
+ \nabla \cdot \boldsymbol{q} - \rho^{s} r^{s} - \rho^{\mathrm{f}} r^{\mathrm{f}} + \left ( \nabla \bar p^{\mathrm{f}} - \rho^{\mathrm{fR}} \boldsymbol{g} \right ) \cdot \boldsymbol{\tilde v}^{\mathrm{f}}_{D} = 0.  
\end{split}
\end{equation}

\subsection{Constitutive equations}
\label{sec:ConstitutiveEquation}
\begin{figure}[!h]
\centering
  \includegraphics[width=0.85\linewidth]{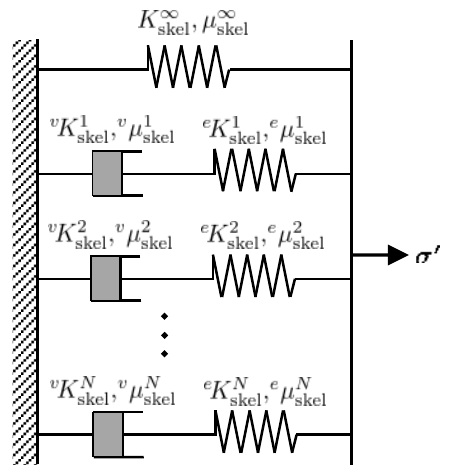}
\caption{Schematic representation of the generalized Maxwell model.}
\label{fig:SchematicGeneralizedMaxwellModel}
\end{figure}

For an isotropic material, the effective stress of the solid skeleton in presence of the volumetric thermal strain can be expressed as:
\begin{equation}\label{eq:LinearViscoConstitutiveEq_1}
\begin{aligned}
\boldsymbol{\sigma'} = 3K_{\mathrm{skel}} \boldsymbol{\varepsilon}^{s,\mathrm{vol}} + 2 \mu_{\mathrm{skel}} \boldsymbol{\varepsilon}^{s,\mathrm{dev}} - 3K_{\mathrm{skel}} \boldsymbol{\varepsilon}^{s,\mathrm{thm}},
\end{aligned}
\end{equation}
where $\mu_{\mathrm{skel}}$ is the shear modulus of porous medium. The total mechanical strain of the solid skeleton is decomposed into the volumetric, $\boldsymbol{\varepsilon}^{s,\mathrm{vol}}$, deviotoric, $\boldsymbol{\varepsilon}^{s,\mathrm{dev}}$, and thermal, $\boldsymbol{\varepsilon}^{s,\mathrm{thm}}$, components. For the viscoelastic behavior, a generalized Maxwell model shown in Figure \ref{fig:SchematicGeneralizedMaxwellModel} is considered. 
The stress-strain relationship for the viscoelastic behavior of the solid skeleton can be characterized in a form of convolution integral:
\begin{equation}\label{eq:LinearViscoConstitutiveEq_3}
\begin{split}
\boldsymbol{\sigma'} = \int_{-\infty}^{t} G^{\mathrm{vol}}(t - \xi) \boldsymbol{\dot \varepsilon}^{s,\mathrm{vol}} (\xi) d\xi \\
+ \int_{-\infty}^{t} G^{\mathrm{dev}}(t - \xi) \boldsymbol{\dot \varepsilon}^{s,\mathrm{dev}} (\xi) d\xi \\ - \int_{-\infty}^{t} G^{\mathrm{thm}}(t - \xi) \boldsymbol{\dot \varepsilon}^{s,\mathrm{thm}} (\xi) d\xi.
\end{split}
\end{equation}
The relaxation modulus function, $G(t)$, that considers viscoelastic behavior of material across a range of time can be expressed as \cite{SH:06}:
\begin{equation}\label{eq:LinearViscoConstitutiveEq_4}
\begin{split}
\begin{dcases}
G^{\mathrm{thm}} (t) = G^{\mathrm{vol}} (t) = 3 K^{\infty}_{\mathrm{skel}} \left (1 + \sum_{m = 1}^{N_{v}} \omega^{k}_{m} \exp \left ( - \frac{t}{\tau^{k}_{m}} \right )  \right ) \\
G^{\mathrm{dev}} (t) = 2 \mu^{\infty}_{\mathrm{skel}} \left (1 + \sum_{m = 1}^{N_{v}} \omega^{\mu}_{m} \exp \left ( - \frac{t}{\tau^{\mu}_{m}} \right )  \right )
\end{dcases}
\end{split}
,
\end{equation}
where $N_{v}$ is the number of spring-dashpot devices. For the $m$th spring-dashpot device, $\omega_{m}$ and $\tau_{m}$ are defined as follows: 
\begin{equation}\label{eq:LinearViscoConstitutiveEq_4_2}
\begin{split}
\omega^{k}_{m} = \frac{^{e}\!K^{m}_{\mathrm{skel}}}{K^{\infty}_{\mathrm{skel}}}, \quad \omega^{\mu}_{m} = \frac{^{e}\!\mu^{m}_{\mathrm{skel}}}{\mu^{\infty}_{\mathrm{skel}}} \quad \tau^{k}_{m} = \frac{^{v}\!K^{m}_{\mathrm{skel}}}{^{e}\!K^{m}_{\mathrm{skel}}}, \quad \tau^{\mu}_{m} = \frac{^{v}\!\mu^{m}_{\mathrm{skel}}}{^{e}\!\mu^{m}_{\mathrm{skel}}},
\end{split}
\end{equation}
with $^{v}\!\mu^{m}_{\mathrm{skel}}$ and $^{v}\!K^{m}_{\mathrm{skel}}$ being the viscous shear and bulk moduli and $^{e}\!\mu^{m}_{\mathrm{skel}}$ and $^{e}\!K^{m}_{\mathrm{skel}}$ being the elastic shear and bulk moduli of the $m$th viscoelastic device, respectively. Assume all stresses before $t = 0$ are zero, then at time $t^{n+1} \geq 0$, the effective stress of the solid skeleton is expressed as: 
\begin{equation}\label{eq:LinearViscoCnvolutionIntegralTime_7}
\begin{split}
\boldsymbol{\sigma'} (t^{n+1}) = \boldsymbol{\sigma'}^{\infty} (t^{n+1}) +  \sum_{m = 1}^{N_{v}} \omega^{k}_{m} \exp \left ( - \frac{t^{n+1}}{\tau^{k}_{m}} \right ) \boldsymbol{\sigma'}^{\infty (\mathrm{vol})} (0) \\
+ \sum_{m = 1}^{N_{v}} \omega^{\mu}_{m} \exp \left ( - \frac{t^{n+1}}{\tau^{\mu}_{m}} \right ) \boldsymbol{\sigma'}^{\infty (\mathrm{dev})} (0) \\
- \sum_{m = 1}^{N_{v}} \omega^{k}_{m} \exp \left ( - \frac{t^{n+1}}{\tau^{k}_{m}} \right ) \boldsymbol{\sigma'}^{\infty (\mathrm{thm})} (0) \\
+ \sum_{m = 1}^{N_{v}} \boldsymbol{h}^{m (\mathrm{vol})} (t^{n+1}) + \sum_{m = 1}^{N_{v}} \boldsymbol{h}^{m (\mathrm{dev})} (t^{n+1}) - \sum_{m = 1}^{N_{v}} \boldsymbol{h}^{m (\mathrm{thm})} (t^{n+1}),
\end{split}
\end{equation}
where $\boldsymbol{\sigma'}^{\infty}$ is the elastic stress defined in the Eq. \eqref{eq:LinearViscoConstitutiveEq_1} and $\boldsymbol{h}^{m (\mathrm{cmp})} (t^{n+1})$ is stress like Internal State Variables (ISVs) defined as \cite{TPG:70}:
\begin{equation}\label{eq:LinearViscoCnvolutionIntegralTime_4}
\begin{split}
\boldsymbol{h}^{\alpha (\mathrm{cmp})} (t^{n+1}) = \exp \left (-\frac{\Delta t}{\tau^{i}_{m}}  \right ) \boldsymbol{h}^{\alpha (\mathrm{cmp})} (t^{n}) \\
+ \omega^{i}_{m} \tau^{i}_{m} \left (\frac{\boldsymbol{\sigma'}^{\infty(\mathrm{cmp})} (t^{n+1}) - \boldsymbol{\sigma'}^{\infty(\mathrm{cmp})} (t^{n}) }{\Delta t}  \right ) \\
\left ( 1 - \exp \left ( - \frac{\Delta t}{\tau^{i}_{m}} \right ) \right ),
\end{split}
\end{equation}
where $i \in [k,\mu,k]$ for a given $\mathrm{cmp} \in [\mathrm{vol}, \mathrm{dev}, \mathrm{thm}]$, respectively. At any time $t$, the viscoelastic ISVs are recovered through a simple recursive update formulation and by knowing the ISVs from the previous time step.

%================================================================================
%
% Finite element analysis
\section{Finite element analysis
        \label{sec:FiniteElementAnalysis}}
A standard Galerkin finite element method is adapted to present the weak form of the governing equations. The displacement, $\boldsymbol{u}$, the pore fluid pressure, $\bar p^{\mathrm{f}}$, and the temperature, $\theta$ are considered as independent state variables. The weak form of governing equations is constructed by multiplying the strong form with a set of admissible weighting functions and integration over the domain. We seek to find $\boldsymbol{u}(\boldsymbol{x},t) \in \mathscr{L}^{\tilde u}$, $\bar p^{\mathrm{f}} (\boldsymbol{x},t) \in \mathscr{L}^{\bar p^{\mathrm{f}}}$, and $\theta (\boldsymbol{x},t) \in \mathscr{L}^{\theta}$ with $t \in [0, T]$ such that:
\begin{equation}\label{eq:FEMFullResolutionModel_1}
\begin{split}
\begin{dcases}
\mathbf{R}_{u} \colon \int_{\Omega} \nabla  \boldsymbol{w}_{u} \cdot \boldsymbol{\sigma} dv - \int_{\Omega} \rho \boldsymbol{w}_{u} \cdot \boldsymbol{b} dv \\
- \int_{\Gamma_{t}}  \boldsymbol{w}_{u} \boldsymbol{\sigma} \cdot \boldsymbol{n} ds = 0 & \in \Omega \\
\mathbf{R}_{\bar p^{\mathrm{f}}} \colon \int_{\Omega} w_{\bar p^{\mathrm{f}}}  \nabla \cdot \boldsymbol{v}^{s} dv - \int_{\Omega} \nabla  w_{\bar p^{\mathrm{f}}} \cdot \boldsymbol{\tilde v}^{\mathrm{f}}_{D} dv \\
- \int_{\Omega} \Big( 3 n^{s} \bar \alpha^{s}_{\theta} + 3 n^{\mathrm{f}} \bar \alpha^{\mathrm{f}}_{\theta} \Big) w_{\bar p^{\mathrm{f}}}  \frac{D^{s} \theta}{Dt} dv \\
- \int_{\Omega} 3 \bar \alpha^{\mathrm{f}}_{\theta} w_{\bar p^{\mathrm{f}}} \nabla \theta^{\mathrm{f}} \cdot \boldsymbol{\tilde v}^{\mathrm{f}}_{D} dv + \int_{\Gamma_{s}} w_{\bar p^{\mathrm{f}}} \boldsymbol{\tilde v}^{\mathrm{f}}_{D} \cdot \boldsymbol{n} ds = 0 & \in \Omega \\
\mathbf{R}_{\theta} \colon \int_{\Omega} w_{\theta} \Big( \rho^{s} C_{p}^{s} + \rho^{\mathrm{f}} C_{p}^{\mathrm{f}} \Big) \frac{D^{s} \theta}{Dt} dv \\
- \int_{\Omega} w_{\theta} \Big( 3 n^{s} \bar \alpha^{s}_{\theta} \theta + 3 n^{\mathrm{f}} \bar \alpha^{\mathrm{f}}_{\theta} \theta \Big) \frac{D^{s} \bar p^{\mathrm{f}}}{Dt} dv \\
+ \int_{\Omega} w_{\theta} \bar{\alpha}_{\theta}^{s} \theta \frac{D^{s} \boldsymbol{\sigma'}}{Dt} \colon \boldsymbol{I} dv + \int_{\Omega} w_{\theta} \rho^{\mathrm{fR}} C_{p}^{\mathrm{f}} \ \nabla \theta \cdot \boldsymbol{\tilde v}^{\mathrm{f}}_{D} dv \\
- \int_{\Omega} w_{\theta} 3 \bar \alpha^{\mathrm{f}}_{\theta} \theta \ \nabla \bar p^{\mathrm{f}} \cdot \boldsymbol{\tilde v}^{\mathrm{f}}_{D} dv  - \int_{\Omega} \nabla  w_{\theta} \cdot \boldsymbol{q} dv \\
- \int_{\Omega} w_{\theta} \big( \rho^{s} r^{s} + \rho^{\mathrm{f}} r^{\mathrm{f}} \big) dv \\
+ \int_{\Omega} w_{\theta} \boldsymbol{\tilde v}^{\mathrm{f}}_{D} \left ( \nabla \bar p^{\mathrm{f}} - \rho^{\mathrm{fR}} \boldsymbol{g} \right ) dv + \int_{\Gamma_{\theta}} w_{\theta} \boldsymbol{q} \cdot \boldsymbol{n} ds = 0 & \in \Omega
\end{dcases}
\end{split}
,
\end{equation}
where $\mathscr{L}^{u}$, $\mathscr{L}^{\bar p^{\mathrm{f}}}$, $\mathscr{L}^{\theta}$ are the trial solution spaces defined as follows:
\begin{equation}\label{eq:FEMFullResolutionModel_3}
\begin{split}
\mathscr{L}^{u} = \left \{ \boldsymbol{u} \colon \Omega \times [0, T] \mapsto \mathbb{R}^{n}, \boldsymbol{u} \in H^{1}, \boldsymbol{u}(t) = \boldsymbol{\tilde u} (t) \right. \\
\left. \text{on} \ \Gamma_{u}, \boldsymbol{u} (\boldsymbol{x}, 0) = \boldsymbol{u}_{0} (\boldsymbol{x}) \right \}, \\
\mathscr{L}^{\bar p^{\mathrm{f}}} = \left \{ \bar p^{\mathrm{f}} \colon \Omega \times [0, T] \mapsto \mathbb{R}^{n}, \bar p^{\mathrm{f}} \in H^{1}, \bar p^{\mathrm{f}} (t) =  \tilde{\bar p}^{\mathrm{f}} (t) \right. \\
\left. \text{on} \ \Gamma_{\mathrm{s}}, \bar p^{\mathrm{f}} (\boldsymbol{x}, 0) = \bar p^{\mathrm{f}}_{0}(\boldsymbol{x}) \right \}, \\
\mathscr{L}^{\theta} = \left \{ \theta \colon \Omega \times [0, T] \mapsto \mathbb{R}^{n}, \theta \in H^{1}, \theta (t) =  \tilde \theta (t) \right. \\
\left. \text{on} \ \Gamma_{\theta}, \theta (\boldsymbol{x}, 0) = \theta_{0} (\boldsymbol{x}) \right \},
\end{split}
\end{equation}
and $\boldsymbol{w}_{u}$, $w_{\bar p^{\mathrm{f}}}$, and $w_{\theta}$ are weighting functions of the independent state variables. $H^{1}$ is the first Sobolev space \cite{Hughes:12}. The symbol $\Omega$ is the volume occupied by the mixture, $\Gamma_{t}$ is the applied traction boundary, $\Gamma_{u}$ is the prescribed displacement boundary, $\Gamma_{s}$ is the prescribed pressure boundary, and $\Gamma_{\theta}$ is the prescribed temperature boundary. The prescribed and initial conditions for the state variables are are denoted by [$\boldsymbol{\tilde u} (t)$, $\tilde{\bar p}^{\mathrm{f}} (t)$, $\tilde \theta (t)$] and [$\boldsymbol{u}_{0} (\boldsymbol{x})$, $\bar p^{\mathrm{f}}_{0} (\boldsymbol{x})$, $\theta_{0} (\boldsymbol{x})$], respectively.

The domain, $\Omega$ is discretized into the non-overlapping sub-domains $\Omega_{e}$, $1 \leq e \leq n_{el}$, where $n_{el}$ is the number of elements in the domain. To avoid oscillation in mechanical stresses, pressure, and temperature, the solutions are approximated by using mixed elements such that a quadratic interpolation is used for the displacement field and linear interpolation is used for approximating pressure and temperature fields. For poro- elastic and viscoelastic problems, numerical oscillations may occur if the spatial and the temporal discretization parameters are not properly chosen \cite{ML:92,Wan:03,PP:11}. To ensure the stability of the system, the stability condition is computed through the relationship given by \cite{FGK:13}. This stability condition depends on the spatial and temporal discretization
step sizes and mechanical parameters of the model. For more information, the reader is referred to \cite{FGK:13}.
%

%================================================================================
%
% Time integration
\subsection{Time integration
\label{sec:FETimeIntegration}}
The weak form of the governing equations is discretized in time by an implicit backward Euler scheme. The discretized governing equations are written in the following compact form:
\begin{equation}\label{eq:FEMFullResolutionModelTimeIntg_3}
\begin{split}
\mathbf{R}^{n+1}_{\mathrm{dyn}} &= \frac{1}{\Delta t} \mathbf{M}^{n+1} \left ( \mathbf{\hat s}^{n+1} -  \mathbf{\hat s}^{n} \right ) + \mathbf{R}^{n+1} (\mathbf{\hat s}^{n+1}) = \mathbf{0}, \\
\mathbf{J}^{n+1}_{\mathrm{dyn}} &= \frac{1}{\Delta t} \mathbf{M}^{n+1} + \mathbf{J}^{n+1} (\mathbf{\hat s}^{n+1}),
\end{split}
\end{equation}
where $\mathbf{R}^{n+1}_{\mathrm{dyn}}$ is the dynamic residual of all governing equations given in Eq.~\eqref{eq:FEMFullResolutionModel_1} and written in a compact form as follows:
\begin{equation}\label{eq:FEMFullResolutionModelTimeIntg_4}
\begin{split}
\mathbf{R}^{n+1}_{\mathrm{dyn}} = \Big[ \mathbf{R}_{u}, \mathbf{R}_{\bar p^{\mathrm{f}}}, \mathbf{R}_{\theta} \Big]^{T}.
\end{split}
\end{equation}
$\mathbf{M}^{n+1}$ is the capacitance matrix that collects the contributions of the inertia terms in Eq.~\eqref{eq:FEMFullResolutionModel_1}. $\mathbf{\hat s}^{n+1}$ and $\mathbf{\hat s}^{n}$ are the vector of all state variables at current and previous time steps, respectively. $\mathbf{R}^{n+1}$ and $\mathbf{J}^{n+1}$ are the residual and jacobian of the static contribution to the dynamic residual, $\mathbf{J}^{n+1}_{\mathrm{dyn}}$ is the analytical derivatives of dynamic residual equations, and $\Delta t$ is the size of time step.
%

%================================================================================
%
% Results and discussions
\section{Results and discussion
        \label{sec:ResultsAndDiscussion}}
In this section, several examples are presented and discussed to illustrate the efficiency of the numerical model. For this purpose, the accuracy of the finite element method is verified through comparison against analytical reference solutions. To gain insight into the effects of different parameters on the performance of articular cartilage, confined and unconfined models are considered. In particular, the effects of material type, temperature, and boundary conditions are investigated. To determine the physical response appropriate for 3D domain, the plain strain assumption is used in 2D models. For all numerical examples, mesh refinement studies are performed and only results for sufficiently fine meshes with negligible discretization errors are shown. Newton-Raphson method is used to solve nonlinear equations using analytically derived jacobians. The discretized linear problem is considered converged if the relative residuals are less than $10^{-9}$. 
 
\subsection{Verification example
        \label{sec:VerificationExamples}}
\begin{figure}[!h]
\centering
  \includegraphics[width=0.7\linewidth]{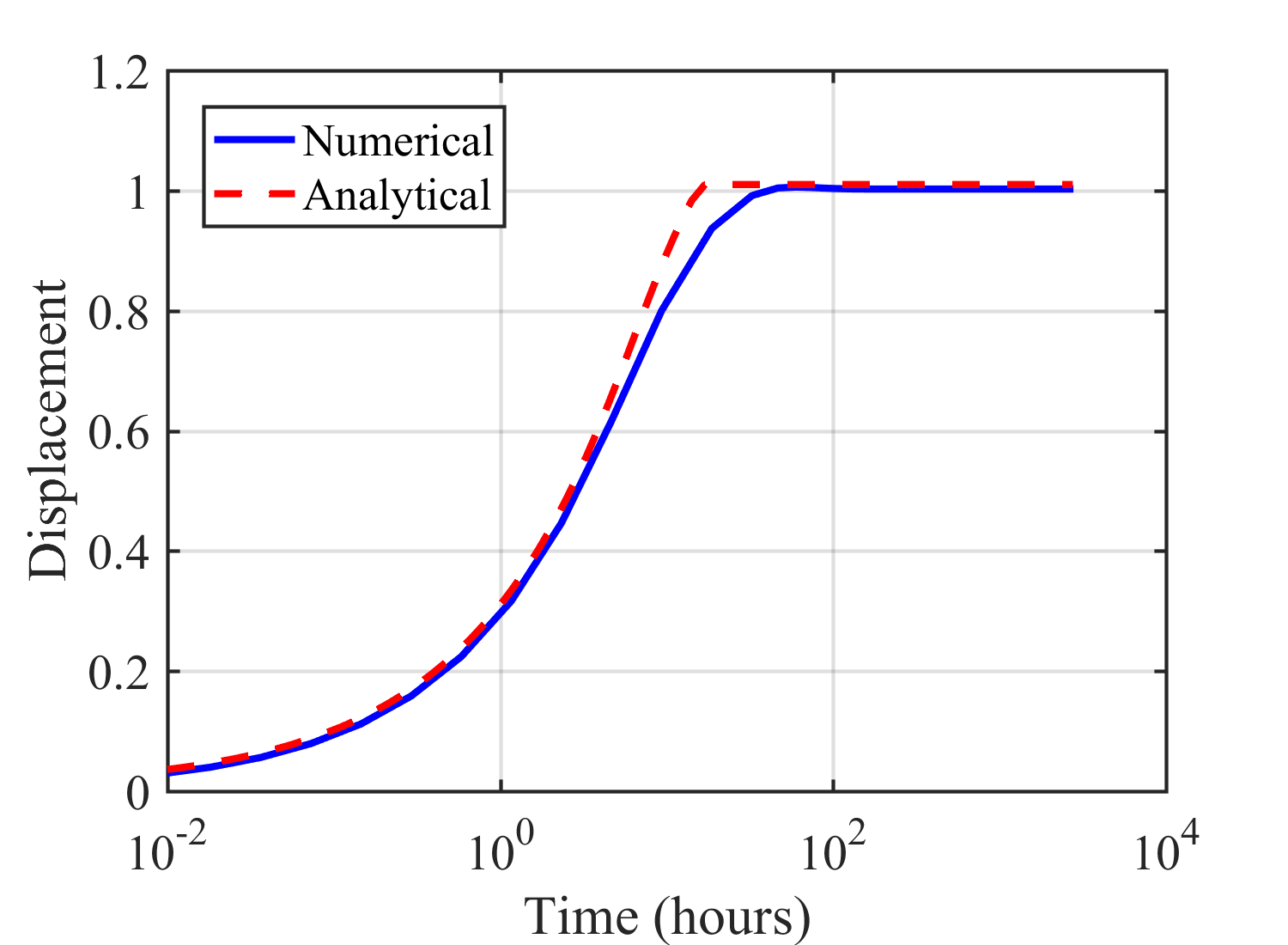}
\caption{Normalized displacement at the top of the model.}
\label{fig:top_node_displacement_thermo_elast_FEM_Bai_withWangParams}
\end{figure}

\begin{figure}[!h]
\centering
  \includegraphics[width=0.70\linewidth]{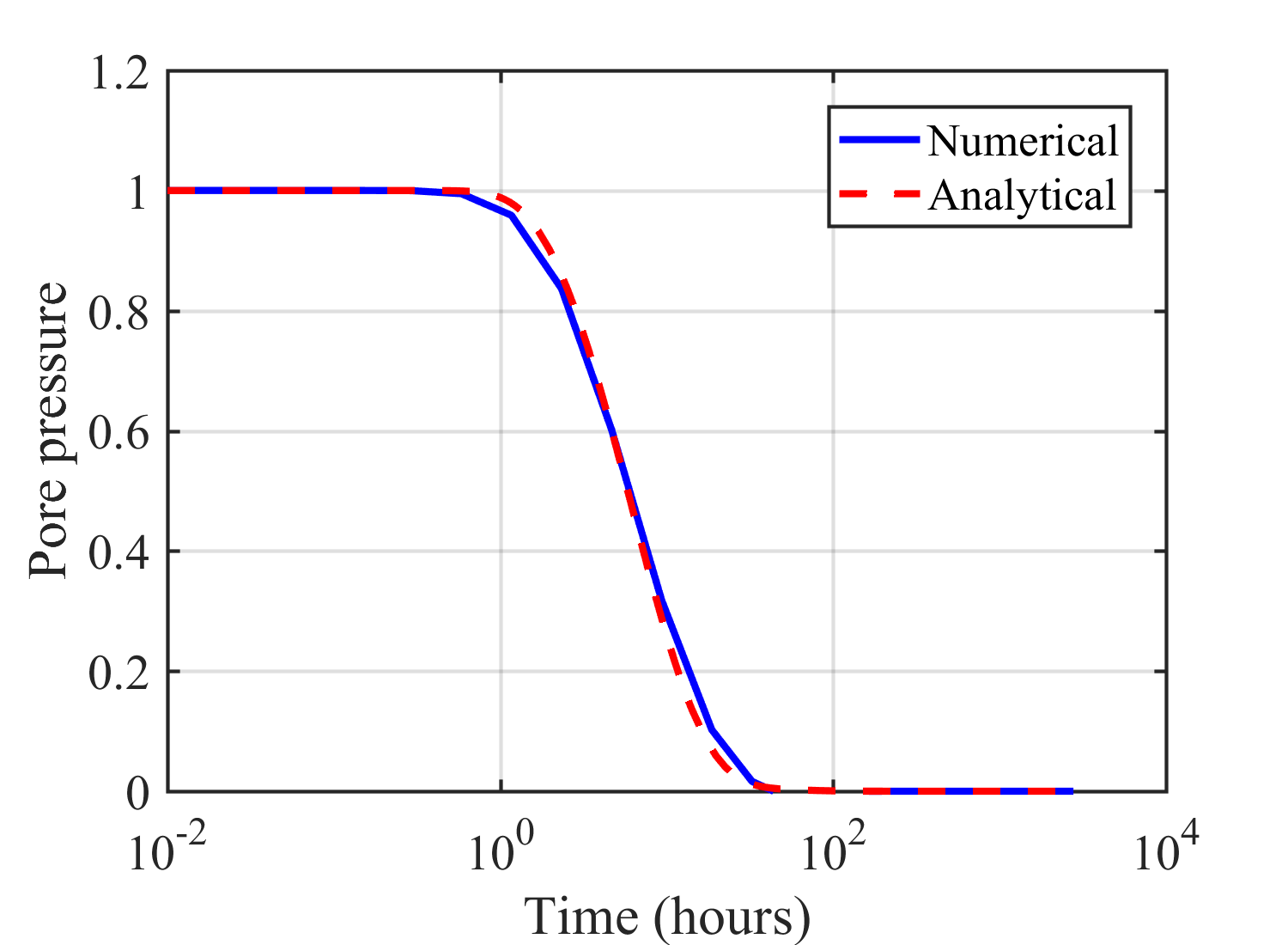}
\caption{Normalized pressure at the bottom of the model.}
\label{fig:bottom_node_pressure_thermo_elast_FEM_Bai_withWangParams}
\end{figure}

\begin{figure}[!h]
\centering
  \includegraphics[width=0.70\linewidth]{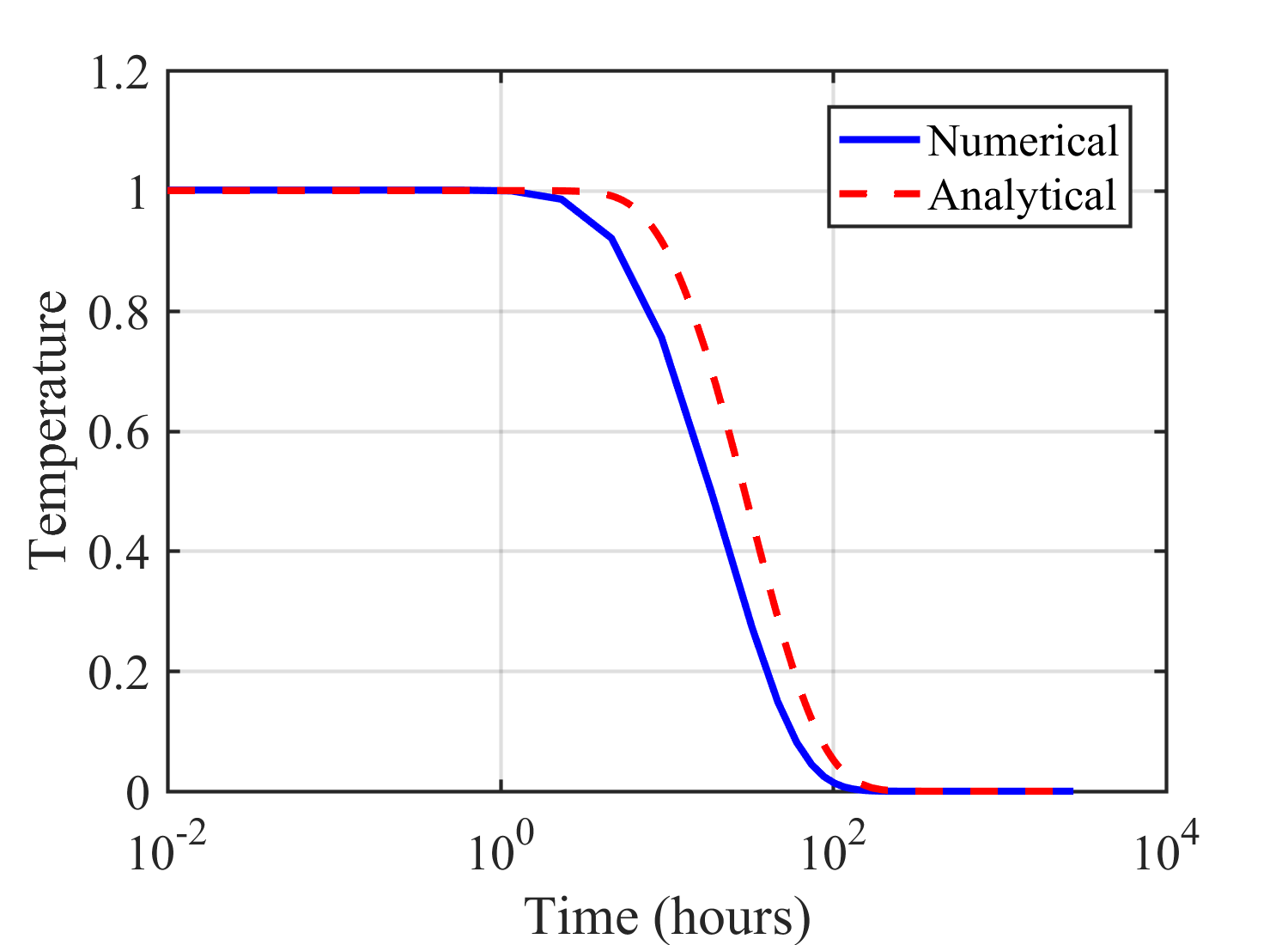}
\caption{Normalized temperature at the bottom of the model.}
\label{fig:bottom_node_temperature_thermo_elast_FEM_Bai_withWangParams}
\end{figure}

The accuracy of the implemented model discussed previously is verified
through comparison against analytical reference solutions. Originally verified for a poro-elastic model \cite{BHR:17}, this study is extended to verify the thermo-poro-elastic response of the model. For this purpose, a laterally confined 1D column with the height $H_{c}$ proposed by Bai et al. \cite{BA:97} is considered. A load with magnitude of $F_{x}$ is applied at the top of the column (i.e at $x = 0$) while the bottom of the column is rigid. At the top of the column, the pressure and temperature are set to zero, which allows the fluid and the temperature to dissipate from the top. The geometric configuration and material parameters are given in Table \ref{tab:1DAnalyticalModelandMaterialProperties}. For numerical modeling, the column is discretized with 100 quadratic elements, and the simulation is continued until the pressure and temperature dissipate from the entire column. The variations of displacement at the top and the pressure and temperature at the bottom of the column are presented as a function of consolidation time as shown in Figures \ref{fig:top_node_displacement_thermo_elast_FEM_Bai_withWangParams}-\ref{fig:bottom_node_temperature_thermo_elast_FEM_Bai_withWangParams}. The displacement, pressure, and temperature are normalized with respect to their maximum values. The results are compared against analytical solutions \cite{BA:97}. The comparison demonstrates good agreement between numerical model and analytical solutions. The small discrepancy appeared in the temperature profile (Figure \ref{fig:bottom_node_temperature_thermo_elast_FEM_Bai_withWangParams}) are caused by  neglecting the forced thermal convection effects in the analytical solutions.
\begin{table*}[!ht]
\begin{center}
\caption{Material properties and model parameters for the laterally confined 1D column.}
\label{tab:1DAnalyticalModelandMaterialProperties}      
\begin{tabular}{llll}
\hline\noalign{\smallskip}
Description & Symbol & Value & Unit \\
\noalign{\smallskip}\hline\noalign{\smallskip}
model height & $H_{c}$ & $0.3$ & $ \mathrm{m} $ \\
fluid volume fraction & $n^{\mathrm{f}}$ & $0.42$ & - \\
hydraulic permeability & $k_{p}$ & $1.0 \times 10^{-14}$ & $ \mathrm{m}^{2}/(\mathrm{Pa \cdot s}) $ \\
Young's modulus of solid skeleton & $K_{\mathrm{skel}}$ & $101.8$ & $ \mathrm{MPa} $ \\
shear modulus of solid skeleton & $\mu_{\mathrm{skel}}$ & $47$ & $ \mathrm{MPa} $ \\
applied traction & $F_{x}$ & $10$ & $ \mathrm{MPa} $ \\
initial pressure & $\bar p^{\mathrm{f}}_{0}$ & $10$ & $ \mathrm{MPa} $ \\
initial temperature & $\theta_{0}$ & $50$ & $ \mathrm{^{o}C} $ \\
real mass density of solid & $\rho^{sR}$ & $3696$ & $ \mathrm{kg}/\mathrm{m}^{3} $ \\
real mass density of fluid & $\rho^{\mathrm{f} R}$ & $1000$ & $ \mathrm{kg}/\mathrm{m}^{3} $ \\
thermal conductivity of solid & $\kappa_{\theta}^{s}$ & $1.38$ & $ \mathrm{W}/(\mathrm{m.K}) $ \\
thermal conductivity of fluid & $\kappa_{\theta}^{\mathrm{f}}$ & $0.6$ & $ \mathrm{W}/(\mathrm{m.K}) $ \\
specific heat capacity of solid & $C_{p}^{s}$ & $703$ & $ \mathrm{J}/(\mathrm{kg.K}) $ \\
specific heat capacity of fluid & $C_{p}^{\mathrm{f}}$ & $4.18 \times 10^{3}$ & $ \mathrm{J}/(\mathrm{kg.K}) $ \\
thermal expansion coefficient of solid & $\bar \alpha_{\theta}^{s}$ & $1.65 \times 10^{-6}$ & $ 1/\mathrm{K} $ \\
thermal expansion coefficient of fluid & $\bar \alpha_{\theta}^{\mathrm{f}}$ & $2.07 \times 10^{-4}$ & $ 1/\mathrm{K} $ \\
\noalign{\smallskip}\hline
\end{tabular}
\end{center}
\end{table*}

\subsection{Confined and unconfined articular cartilage models
        \label{sec:ArticularCartilageModelConUnCon}}
\begin{figure}[!h]
\centering
\subfloat[confined compression]{
  \includegraphics[width=0.49\linewidth]{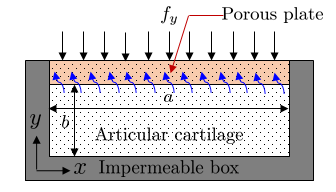}
}
\subfloat[unconfined compression]{
  \includegraphics[width=0.49\linewidth]{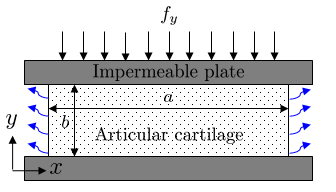}
}
\caption{Schematic of confined and unconfined compression models.}
\label{fig:ArticularCartilageConfinedUnconfinedSchematic}
\end{figure}

To explore the characteristics of the articular cartilage model, both confined and unconfined boundary conditions are considered, as shown in Figure \ref{fig:ArticularCartilageConfinedUnconfinedSchematic}. For the confined model (Figure~\ref{fig:ArticularCartilageConfinedUnconfinedSchematic}~(a)), articular cartilage is compressed into a perfectly smooth and impermeable box. The fluid pore pressure and temperature can dissipate from the top boundary. The model is fixed at the bottom and displacements in the $x$ direction are constrained to zero along the sides. A porous plate at the top of the model is used to uniformly transfer the  applied load, $f_{y}$, to the articular cartilage specimen. For numerical modeling, the domain is discretized with $55 \times 110$ elements; the geometric configuration and material parameters, used in the test, are shown in Table \ref{tab:ArticularCartilageModelandMaterialProperties}.  
\begin{table*}[!ht]
\begin{center}
\caption{Material properties and model parameters for the articular cartilage model \cite{HA:09, ALM:84, LS:17}.}
\label{tab:ArticularCartilageModelandMaterialProperties}      
\begin{tabular}{llll}
\hline\noalign{\smallskip}
Description & Symbol & Value & Unit \\
\noalign{\smallskip}\hline\noalign{\smallskip}
specimen wide & $a$ & $0.25 \times 10^{-3}$ & $ \mathrm{m} $ \\
specimen thickness & $b$ & $0.50 \times 10^{-3}$ & $ \mathrm{m} $ \\
fluid volume fraction & $n^{\mathrm{f}}$ & $0.8$ & - \\
hydraulic permeability & $k_{p}$ & $1.0 \times 10^{-15}$ & $ \mathrm{m}^{2}/(\mathrm{Pa \cdot s}) $ \\
bulk modulus of solid skeleton & $K_{\mathrm{skel}}$ & $0.2$ & $ \mathrm{MPa} $ \\
shear modulus of solid skeleton & $\mu_{\mathrm{skel}}$ & $0.1$ & $ \mathrm{MPa} $ \\
number of spring-dashpot devices & $N_{v}$ & $5$ & - \\
elastic bulk moduli of devices & $^{e}\!K^{m}_{\mathrm{skel}}$ & $[89, 36, 36, 8.9, 2.9]$ & $ \mathrm{KPa} $ \\
elastic shear moduli of devices & $^{e}\!\mu^{m}_{\mathrm{skel}}$ & $[97, 27, 16, 9.7, 1.7]$ & $ \mathrm{KPa} $ \\
viscous bulk moduli of devices & $^{v}\!K^{m}_{\mathrm{skel}}$ & $[0.089, 0.036, 3.6, 8.9, 29]$ & $ \mathrm{KPa} $ \\
viscous shear moduli of devices & $^{v}\!\mu^{m}_{\mathrm{skel}}$ & $[0.097, 0.27, 1.6, 9.7, 17]$ & $ \mathrm{KPa} $ \\
applied traction & $f_{y}$ & $5.0 \times 10^{4}$ & $ \mathrm{Pa} $ \\
real mass density of solid & $\rho^{sR}$ & $1260$ & $ \mathrm{kg}/\mathrm{m}^{3} $ \\
real mass density of fluid & $\rho^{\mathrm{f} R}$ & $997.1$ & $ \mathrm{kg}/\mathrm{m}^{3} $ \\
thermal conductivity of solid & $\kappa_{\theta}^{s}$ & $0.6$ & $ \mathrm{W}/(\mathrm{m.K}) $ \\
thermal conductivity of fluid & $\kappa_{\theta}^{\mathrm{f}}$ & $0.6$ & $ \mathrm{W}/(\mathrm{m.K}) $ \\
specific heat capacity of solid & $C_{p}^{s}$ & $3.17 \times 10^{3}$ & $ \mathrm{J}/(\mathrm{kg.K}) $ \\
specific heat capacity of fluid & $C_{p}^{\mathrm{f}}$ & $4.18 \times 10^{3}$ & $ \mathrm{J}/(\mathrm{kg.K}) $ \\
thermal expansion coefficient of solid & $\bar \alpha_{\theta}^{s}$ & $0.01 \times 10^{-3}$ & $ 1/\mathrm{K} $ \\
thermal expansion coefficient of fluid & $\bar \alpha_{\theta}^{\mathrm{f}} @ 24 \ ^{\mathrm{o}}\mathrm{C}$ & $0.246 \times 10^{-3}$ & $ 1/\mathrm{K} $ \\
thermal expansion coefficient of fluid & $\bar \alpha_{\theta}^{\mathrm{f}} @ 37.5 \ ^{\mathrm{o}}\mathrm{C}$ & $0.376 \times 10^{-3}$ & $ 1/\mathrm{K}$ \\
thermal expansion coefficient of fluid & $\bar \alpha_{\theta}^{\mathrm{f}} @ 55 \ ^{\mathrm{o}}\mathrm{C}$ & $0.544 \times 10^{-3}$ & $ 1/\mathrm{K}$ \\
\noalign{\smallskip}\hline
\end{tabular}
\end{center}
\end{table*}

\begin{figure}[!h]
\centering
\subfloat[displacement]{
  \includegraphics[width=0.49\linewidth]{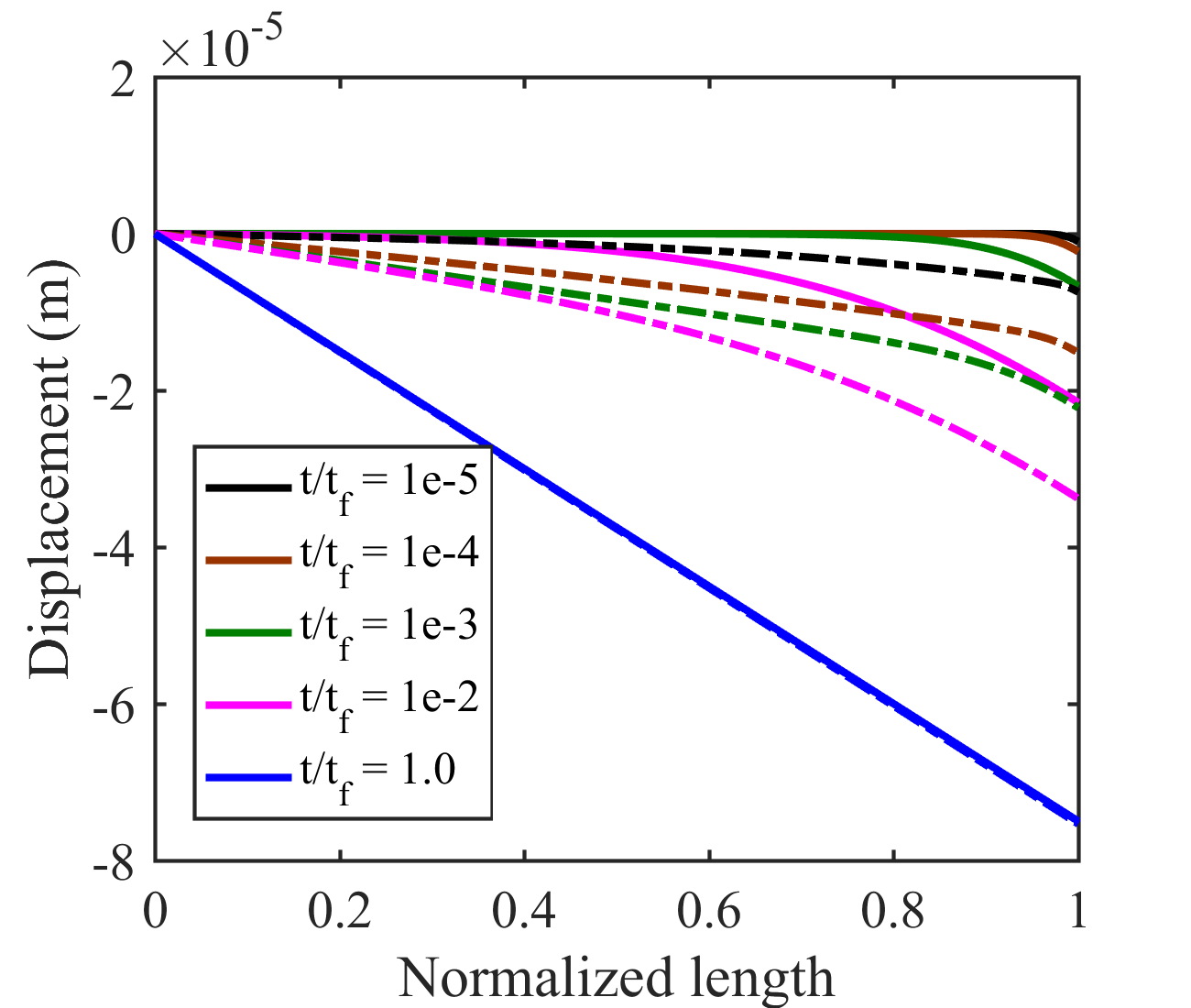}
}
\subfloat[pore fluid pressure]{
  \includegraphics[width=0.49\linewidth]{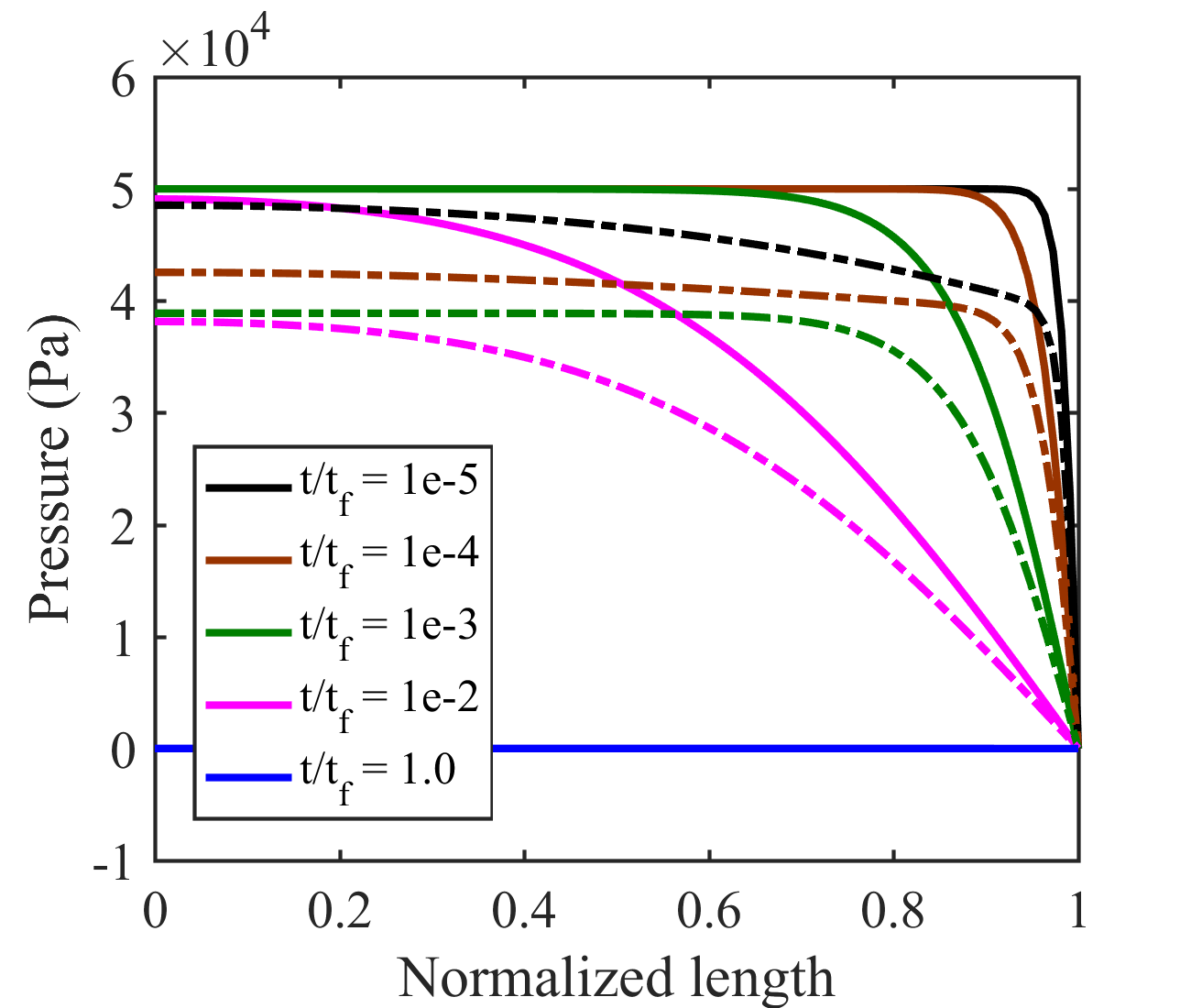}
}
\caption{Variations of displacement and pore fluid pressure along the side of the confined model and for different loading time. The solid lines represent the poro-elastic responses without thermal coupling and the dashed lines show the thermo-poro-elastic responses.}
\label{fig:xmin_displacement_pressure_poroelast_thermoporoelast_37_5C}
\end{figure}

\begin{figure}[!h]
\centering
\subfloat[displacement]{
  \includegraphics[width=0.49\linewidth]{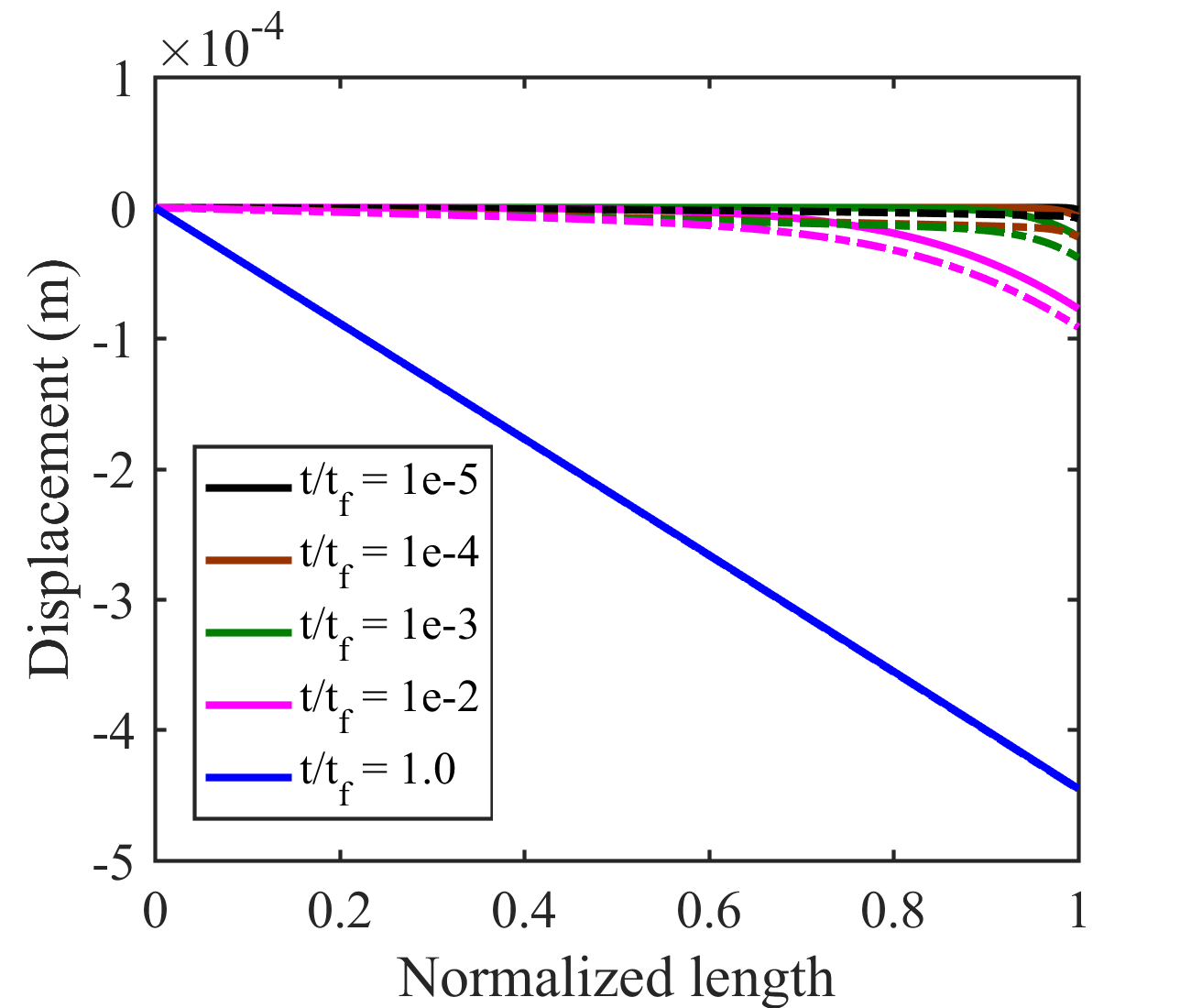}
}
\subfloat[pore fluid pressure]{
  \includegraphics[width=0.49\linewidth]{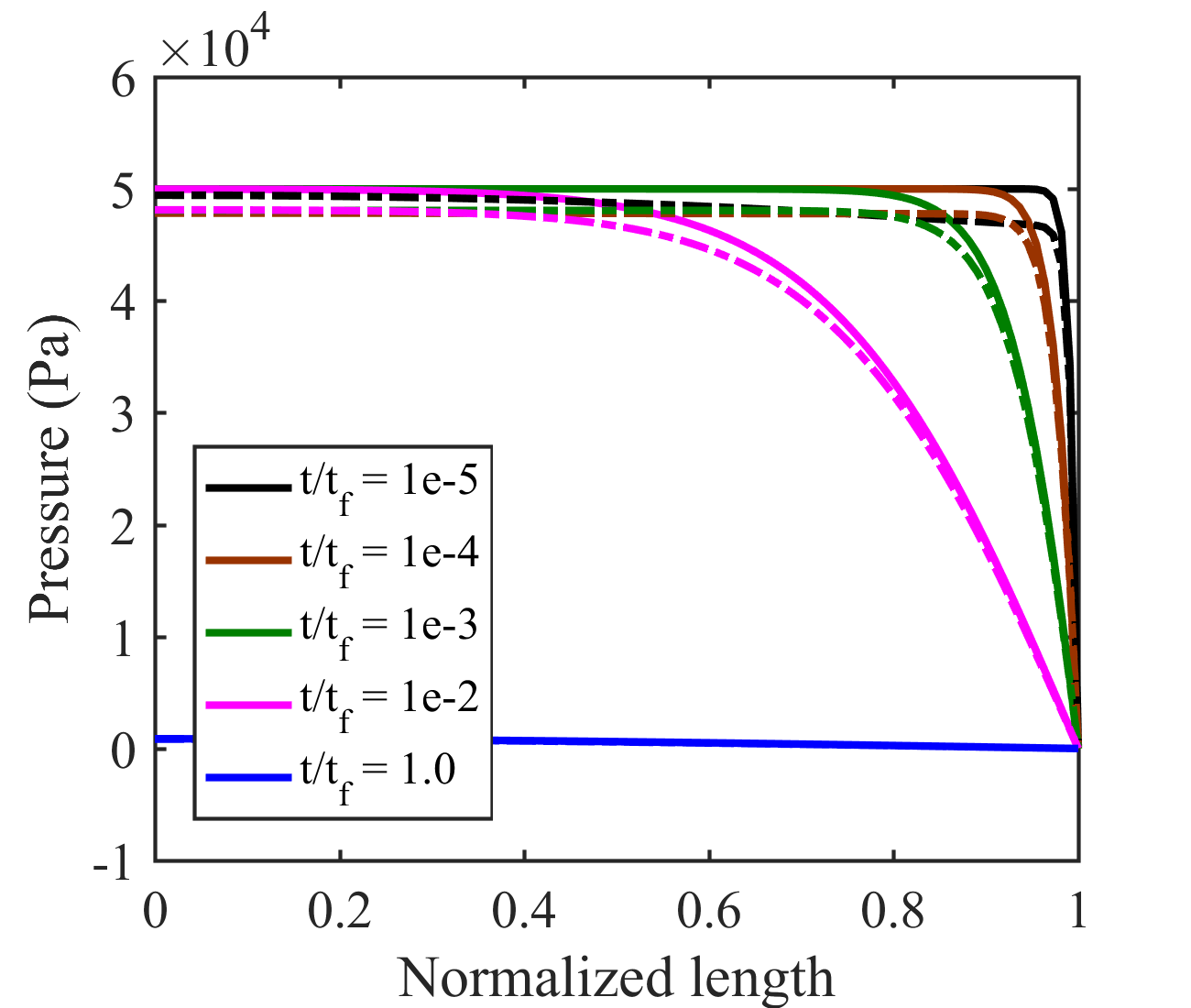}
}
\caption{Variations of displacement and pore fluid pressure along the side of the confined model and for different loading time. The solid lines represent the poro-viscoelastic responses without thermal coupling and the dashed lines show the thermo-poro-viscoelastic responses.}
\label{fig:xmin_displacement_pressure_poroviscoelast_thermoporoviscoelast_37_5C}
\end{figure}

The behavior of the articular cartilage model is studied considering poro- elastic and viscoelastic material behaviors. The results are compared with those that consider the influence of temperature on both poro- elastic and viscoelastic behavior through thermal coupling. To this end, the confined model is analyzed at $\theta = 37.5 ^{\text{o}} \text{C}$. The results are presented in terms of variations of displacement and pore fluid pressure along the side of the confined model and for different loading times. The final loading time, $t_{\mathrm{f}}$, that allows full dissipation of pore fluid pressure and temperature in the model is set to $4 \times 10^{3}$ seconds. For poro- elastic and viscoelastic materials, the displacement and pore fluid pressure profiles are shown in Figures \ref{fig:xmin_displacement_pressure_poroelast_thermoporoelast_37_5C} and \ref{fig:xmin_displacement_pressure_poroviscoelast_thermoporoviscoelast_37_5C} , respectively, where solid lines represent responses without thermal coupling and dashed lines show those with thermal coupling. The results of poro- elastic and viscoelastic models show that at very early times, large fluid pore pressure results in small displacements. As the pore fluid pressure dissipates, the deformation increases until a purely elastic response is recovered. For the poro-elastic behavior, models with thermal effects show a significant reduction in both pore fluid pressure and displacement at early times. However, for the poro-viscoelastic behavior, this reduction is not considerable due to creep recovery and constant stress over the loading period.

\begin{figure}[!h]
\centering
\subfloat[displacement]{
  \includegraphics[width=0.49\linewidth]{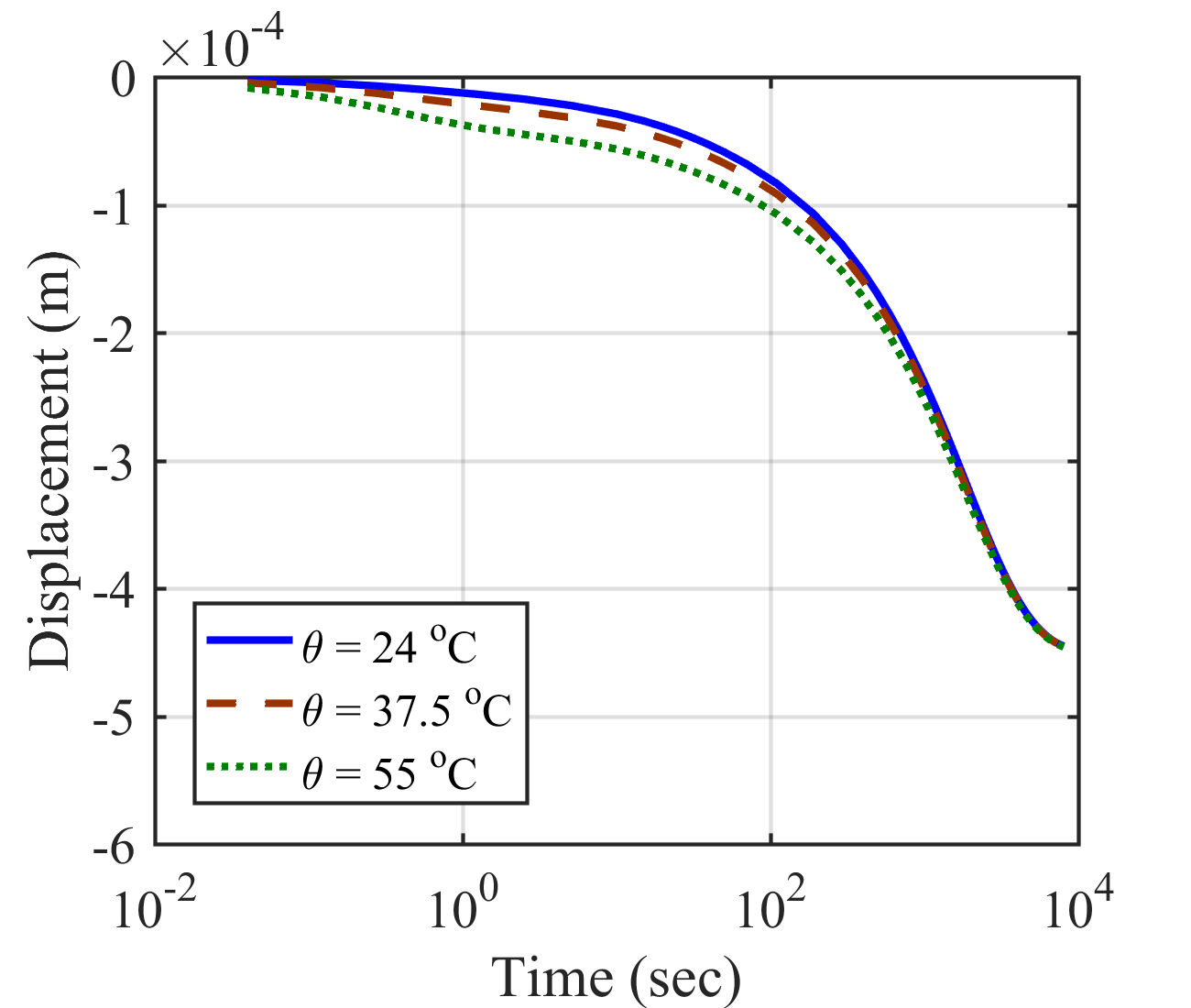}
}
\subfloat[pore fluid pressure]{
  \includegraphics[width=0.49\linewidth]{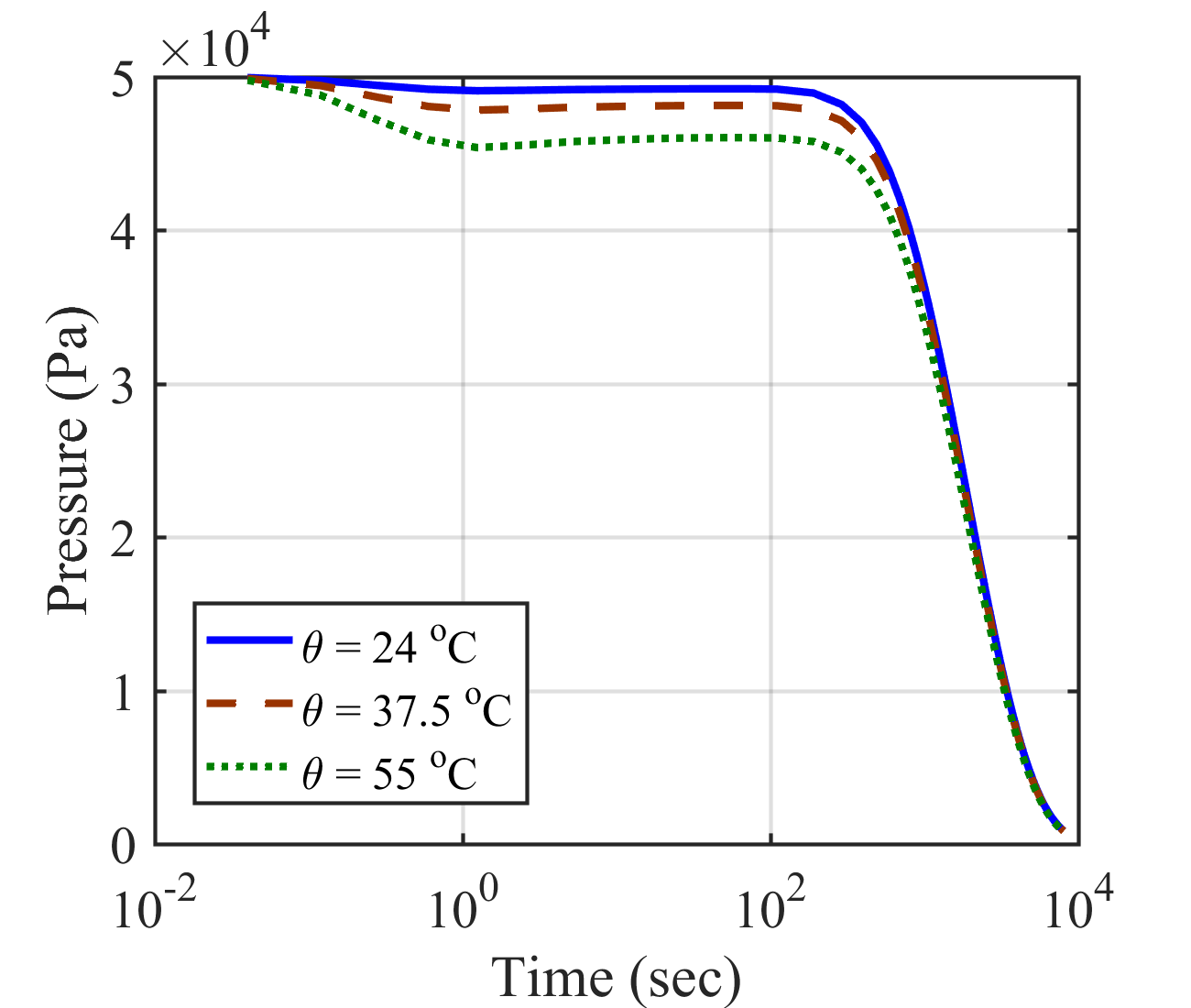}
}
\caption{Evolution of $y$-displacement and pore fluid pressure at the top and bottom midpoints of the confined model and for different temperatures.}
\label{fig:displacement_pressure_thermoporoviscoelast_confined_24_375_55C}
\end{figure}

\begin{figure}[!h]
\centering
\subfloat[displacement]{
  \includegraphics[width=0.49\linewidth]{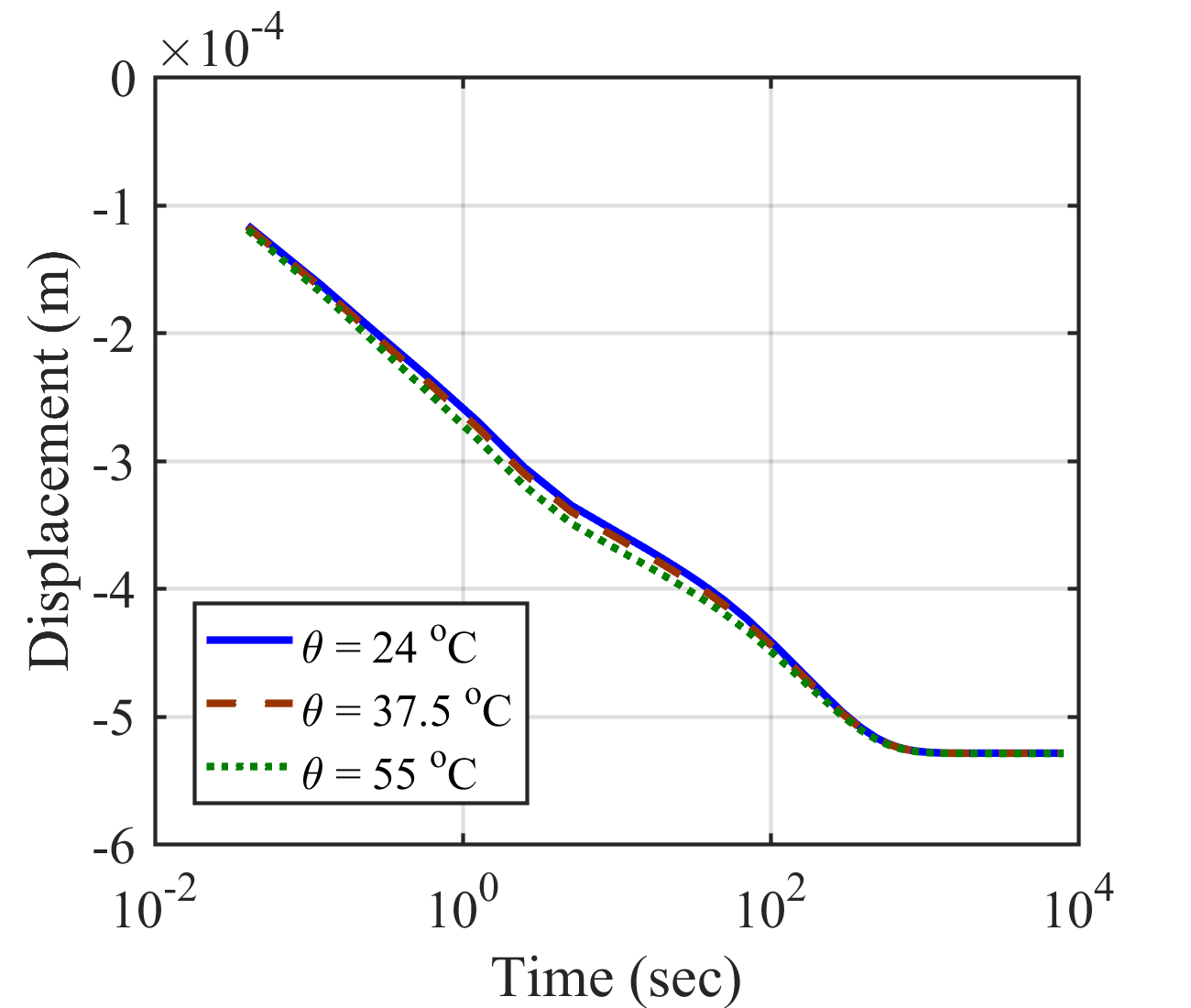}
}
\subfloat[pore fluid pressure]{
  \includegraphics[width=0.49\linewidth]{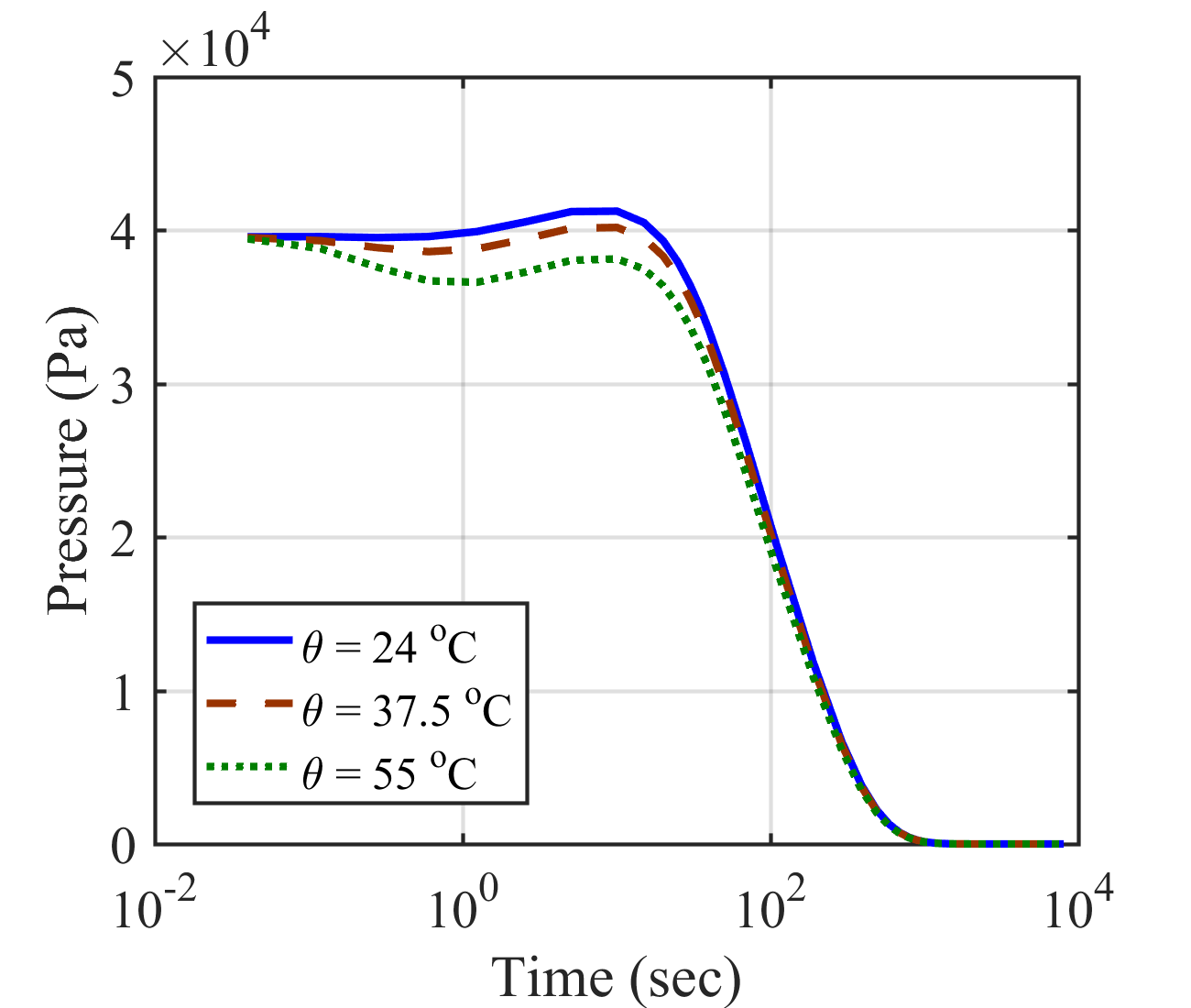}
}
\caption{Evolution of $y$-displacement and pore fluid pressure at the top and bottom midpoints of the unconfined model and for different temperatures.}
\label{fig:displacement_pressure_thermoporoviscoelast_unconfined_24_375_55C}
\end{figure}

In order to explore the effect of temperature on the behavior of articular cartilage, different temperatures ($24, 37.5, \text{and} \ 55 ^{\text{o}} \text{C}$) are considered for both confined and unconfined models with thermo-poro-viscoelastic behavior. For the unconfined model, as shown in Figure \ref{fig:ArticularCartilageConfinedUnconfinedSchematic} (b), the fluid pore pressure can dissipate along the sides, and the temperature can dissipate from the top of the model. At the bottom, the displacements in both $x$ and $y$ directions are constrained to zero. Impermeable solid plates at the top and bottom of the model are used. The top plate uniformly transfers the applied load to the articular cartilage specimen. For both confined and unconfined models, the evolution of $y$-displacement at the top midpoint and pore fluid pressure at the bottom midpoint are presented for different temperatures, as shown in Figures \ref{fig:displacement_pressure_thermoporoviscoelast_confined_24_375_55C} and \ref{fig:displacement_pressure_thermoporoviscoelast_unconfined_24_375_55C}, respectively. The results reveal the influence of temperature on the articular cartilage behavior. The results show a significant influence of temperature at very early times. After dissipation of pore fluid pressure and temperature, elastic response is recovered in both confined and unconfined models.

\subsection{Partially loaded confined articular cartilage
        \label{sec:ConfinedArticularCartilagewithPartialLoad}}
\begin{figure}[!h]
\centering
  \includegraphics[width=0.7\linewidth]{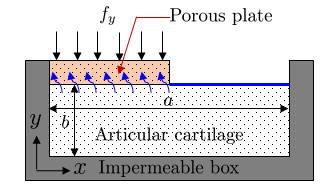}
\caption{Confined compression articular cartilage model with partial loading. Dissipation of pore fluid pressure and temperature is not allowed through the solid blue line.}
\label{fig:ArticularCartilageConfinedHalfLoadedSchematic}
\end{figure}

The effects of partial loading is an extremely important facet of articular cartilage mechanics on account of the fact that medical injuries in articular cartilage are mostly due to partial loads. The following example extends the confined model (Figure~\ref{fig:ArticularCartilageConfinedUnconfinedSchematic}~(a)) to study the behavior of articular cartilage under partial loading condition shown in Figure~\ref{fig:ArticularCartilageConfinedHalfLoadedSchematic}. All model and material parameters are adopted from Table \ref{tab:ArticularCartilageModelandMaterialProperties}, except for the width of the specimen which is set to $a = 5.0 \times 10^{-3} m$. The fluid pore pressure and temperature can dissipate only from the bottom of the porous plate. The displacement in the $x$ direction are constrained to zero along the sides, and all displacements are fixed at the bottom. Again, a porous plate at the top of the model is used to uniformly transfer the applied load, $f_{y}$, to the articular cartilage specimen. The domain is discretized with $125 \times 25$ quadratic elements and the final loading time, $t_{\mathrm{f}}$, that allows full dissipation of pore fluid pressure and temperature is set to $2 \times 10^{5}$ seconds.

The viscoelastic behavior is studied at different temperatures and the evolution of displacement along the top and pore fluid pressure along the bottom of the model are given at different loading stages. Initially, high pore fluid pressure is established in the half of the model where the pressure is allowed to dissipate. However, as the simulation proceeds the vertical displacement increases and the pore fluid pressure dissipates from the bottom of the porous plate. 

\begin{figure}[!h]
\centering
  \includegraphics[width=0.99\linewidth]{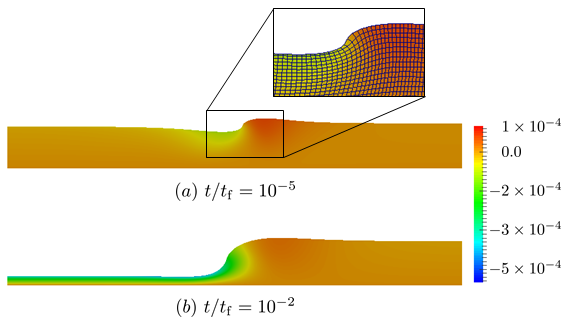}
\caption{Displacement distribution in the articular cartilage specimen at $\theta = 37.5 \mathrm{^oC}$ and for different loading stages.}
\label{fig:confined_half_loaded_left_displacement_t_tf_1en5_1en2}
\end{figure}

\begin{figure}[!h]
\centering
  \includegraphics[width=0.99\linewidth]{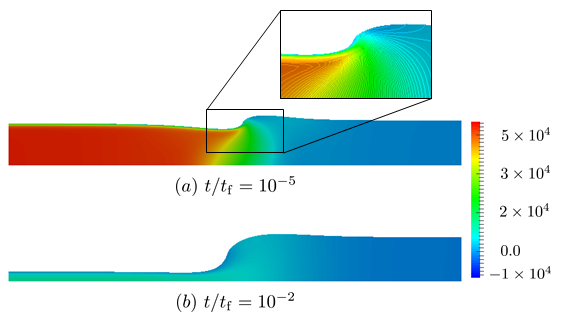}
\caption{Pore fluid pressure distribution in the articular cartilage specimen at $\theta = 37.5 \mathrm{^oC}$ and for different loading stages.}
\label{fig:confined_half_loaded_left_pressure_t_tf_1en5_1en2}
\end{figure}

The results also show that at very early times, temperature has a significant influence on the articular cartilage behavior. High temperature results in low pressures and large vertical displacements. The results also show fast dissipation of temperature at early times.  Again, once the temperature dissipates, the linear elastic response is recovered. The inset snapshots given in Figures \ref{fig:confined_half_loaded_left_displacement_t_tf_1en5_1en2} and \ref{fig:confined_half_loaded_left_pressure_t_tf_1en5_1en2} at the early loading stages depict a significant pressure drop and large displacement in the half of the model. Due to free displacement boundary conditions on the right-half, expansion occurs at the top middle of the model. 

\begin{figure}[!h]
\centering
  \includegraphics[width=0.7\linewidth]{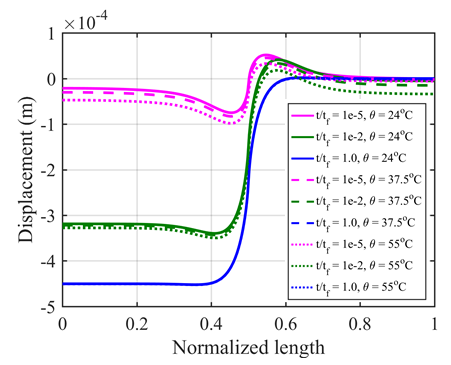}
\caption{Evolution of displacement along the top of the model and for different temperatures and loading stages.}
\label{fig:ymax_displacement_thermoporovisco_half_loaded_left_24_375_55C}
\end{figure}
\begin{figure}[!h]
\centering
  \includegraphics[width=0.7\linewidth]{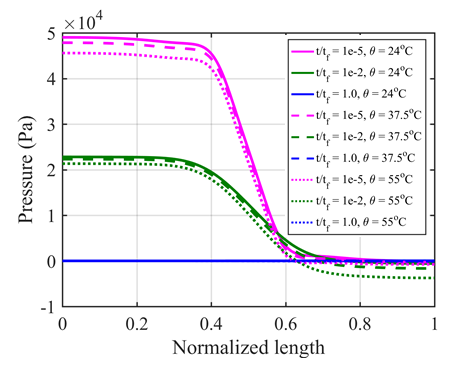}
\caption{Evolution of pore fluid pressure along the bottom of the model and for different temperatures and loading stages.}
\label{fig:ymin_pressure_thermoporovisco_half_loaded_left_24_375_55C}
\end{figure}
%

%================================================================================
%
% Conclusions
\section{Conclusions
        \label{sec:Conclusions}}
This paper presents a finite element methodology for numerical modeling of thermo-poro- elastic and viscoelastic behavior of articular cartilage. The thermo-poro-mechanical behavior of a fully saturated articular cartilage is described by a biphasic solid-fluid mixture theory. Governing and balance equations are considered to describe the displacement in the solid skeleton, fluid flow, and temperature distribution in the mixture. The mechanical behavior of the solid skeleton is described using linear elastic and viscoelastic isotropic materials. Time-dependent response is considered to model the transient thermo-poro-viscoelastic behavior of articular cartilage. The thermal influence on the mixture behavior is described through the temperature dependent mass density and volumetric thermal strain. The Darcy's law is used to describe the fluid flow in the mixture. 

The presented finite element framework proved reliable for a variety of 2D articular cartilage models. The accuracy of the implemented framework is verified through comparison against analytical reference solutions. To gain insight into the characteristics of articular cartilage, both confined and unconfined models are considered. The behavior of articular cartilage is investigated for both elastic and viscoelastic materials, with and without thermal coupling. The influence of temperature is studied through examining the behavior of articular cartilage for different temperatures. Partially loaded confined articular cartilage model is considered to investigate deformation, pore fluid pressure and temperature dissipation processes.   The results show a significant influence of temperature on the behavior of articular cartilage at early loading stages. As the temperature and pore fluid dissipate from the system, the steady-state elastic behavior is recovered.

While only 2D numerical models are considered in this study, the developed method allows for the convenient extension of the framework to 3D behavior of articular cartilage. For future studies this method needs to be extended to explore the 3D characteristics of articular cartilage in the finite strain regime.
%

%================================================================================
%
% Conflict of interest statement
\section*{Conflict of interest statement}
The authors declare that there is no conflict of interest.
%

%================================================================================
%
% Funding sources
\section*{Funding sources}
This research did not receive any specific grant from funding agencies in the public, commercial, or not-for-profit sectors.
%

%================================================================================
%
% References
%\section*{References
%        \label{sec:References}}
%
\bibliographystyle{plain} 
\bibliography{mybibfile} 

\begin{thebibliography}{10}
\expandafter\ifx\csname url\endcsname\relax
  \def\url#1{\texttt{#1}}\fi
\expandafter\ifx\csname urlprefix\endcsname\relax\def\urlprefix{URL }\fi
\expandafter\ifx\csname href\endcsname\relax
  \def\href#1#2{#2} \def\path#1{#1}\fi

\bibitem{MAS:93}
V.~C. Mow, G.~A. Ateshian, R.~L. Spilker, Biomechanics of diarthrodial joints:
  a review of twenty years of progress, Journal of biomechanical engineering
  115~(4B) (1993) 460--467.

\bibitem{FBR:09}
A.~J. Sophia~Fox, A.~Bedi, S.~A. Rodeo, The basic science of articular
  cartilage: structure, composition, and function, Sports health 1~(6) (2009)
  461--468.

\bibitem{PCM+:05}
E.~Pena, B.~Calvo, M.~Martinez, D.~Palanca, M.~Doblar{\'e}, Finite element
  analysis of the effect of meniscal tears and meniscectomies on human knee
  biomechanics, Clinical Biomechanics 20~(5) (2005) 498--507.

\bibitem{Mak:86}
A.~Mak, The apparent viscoelastic behavior of articular cartilage—the
  contributions from the intrinsic matrix viscoelasticity and interstitial
  fluid flows, Journal of biomechanical engineering 108~(2) (1986) 123--130.

\bibitem{SSH:08}
R.~Shirazi, A.~Shirazi-Adl, M.~Hurtig, Role of cartilage collagen fibrils
  networks in knee joint biomechanics under compression, Journal of
  biomechanics 41~(16) (2008) 3340--3348.

\bibitem{GLA+:09}
S.~Gupta, J.~Lin, P.~Ashby, L.~Pruitt, A fiber reinforced poroelastic model of
  nanoindentation of porcine costal cartilage: a combined experimental and
  finite element approach, Journal of the mechanical behavior of biomedical
  materials 2~(4) (2009) 326--338.

\bibitem{GL:11}
K.~Gu, L.~Li, A human knee joint model considering fluid pressure and fiber
  orientation in cartilages and menisci, Medical engineering \& physics 33~(4)
  (2011) 497--503.

\bibitem{HSK+:03}
C.-Y. Huang, M.~A. Soltz, M.~Kopacz, V.~C. Mow, G.~A. Ateshian, et~al.,
  Experimental verification of the roles of intrinsic matrix viscoelasticity
  and tension-compression nonlinearity in the biphasic response of cartilage,
  Transactions-American Society of Mechanical Engineers Journal of
  Biomechanical Engineering 125~(1) (2003) 84--93.

\bibitem{NTJ:03}
P.~Netti, F.~Travascio, R.~Jain, Coupled macromolecular transport and gel
  mechanics: poroviscoelastic approach, AIChE Journal 49~(6) (2003) 1580--1596.

\bibitem{HA:09}
S.~K. Hoang, Y.~N. Abousleiman, Poroviscoelastic two-dimensional anisotropic
  solution with application to articular cartilage testing, Journal of
  engineering mechanics 135~(5) (2009) 367--374.

\bibitem{TGF:14}
A.~Tomic, A.~Grillo, S.~Federico, Poroelastic materials reinforced by
  statistically oriented fibres—numerical implementation and application to
  articular cartilage, The IMA Journal of Applied Mathematics 79~(5) (2014)
  1027--1059.

\bibitem{JF:10}
R.~K. June, D.~P. Fyhrie, Temperature effects in articular cartilage
  biomechanics, Journal of Experimental Biology 213~(22) (2010) 3934--3940.

\bibitem{LS:99}
R.~Lewis, B.~Schrefler, The finite element method in the static and dynamic
  deformation and consolidation of porous media, Meccanica 34~(3) (1999)
  231--232.

\bibitem{Deboer:06}
R.~De~Boer, Trends in continuum mechanics of porous media, Vol.~18, Springer
  Science \& Business Media, 2006.

\bibitem{GSG:96}
D.~Gawin, B.~A. Schrefler, M.~Galindo, Thermo-hydro-mechanical analysis of
  partially saturated porous materials, Engineering Computations 13~(7) (1996)
  113--143.

\bibitem{Holzapfel:00}
G.~A. Holzapfel, Nonlinear solid mechanics, Vol.~24, Wiley Chichester, 2000.

\bibitem{CN:63}
B.~D. Coleman, W.~Noll, The thermodynamics of elastic materials with heat
  conduction and viscosity, Archive for Rational Mechanics and Analysis 13~(1)
  (1963) 167--178.

\bibitem{Coussy:04}
O.~Coussy, Poromechanics, John Wiley \& Sons, 2004.

\bibitem{SH:06}
J.~C. Simo, T.~J. Hughes, Computational inelasticity, Vol.~7, Springer Science
  \& Business Media, 2006.

\bibitem{TPG:70}
R.~L. Taylor, K.~S. Pister, G.~L. Goudreau, Thermomechanical analysis of
  viscoelastic solids, International Journal for Numerical Methods in
  Engineering 2~(1) (1970) 45--59.

\bibitem{Hughes:12}
T.~J. Hughes, The finite element method: linear static and dynamic finite
  element analysis, Courier Corporation, 2012.

\bibitem{ML:92}
M.~A. Murad, A.~F. Loula, Improved accuracy in finite element analysis of
  biot's consolidation problem, Computer Methods in Applied Mechanics and
  Engineering 95~(3) (1992) 359--382.

\bibitem{Wan:03}
J.~Wan, Stabilized finite element methods for coupled geomechanics and
  multiphase flow, Ph.D. thesis, Stanford university (2003).

\bibitem{PP:11}
M.~Preisig, J.~H. Pr{\'e}vost, Stabilization procedures in coupled
  poromechanics problems: A critical assessment, International Journal for
  Numerical and Analytical Methods in Geomechanics 35~(11) (2011) 1207--1225.

\bibitem{FGK:13}
M.~Favino, A.~Grillo, R.~Krause, A stability condition for the numerical
  simulation of poroelastic systems, in: Poromechanics V: Proceedings of the
  Fifth Biot Conference on Poromechanics, 2013, pp. 919--928.

\bibitem{BHR:17}
R.~Behrou, A thermo-poro-viscoelastic model for the behavior of polymeric
  porous separators, ECS Transactions 80~(8) (2017) 583--591.

\bibitem{BA:97}
M.~Bai, Y.~Abousleiman, Thermoporoelastic coupling with application to
  consolidation, International Journal for Numerical and Analytical Methods in
  Geomechanics 21~(2) (1997) 121--132.

\bibitem{ALM:84}
C.~Armstrong, W.~Lai, V.~Mow, An analysis of the unconfined compression of
  articular cartilage, Journal of biomechanical engineering 106~(2) (1984)
  165--173.

\bibitem{LS:17}
B.~Loret, F.~M.~F. Simoes, Biomechanical aspects of soft tissues, Crc Press,
  2017.

\end{thebibliography}
%

%================================================================================

\end{document}